\documentclass[aps,prd,preprint,superscriptaddress,showpacs,preprintnumbers,amsmath,amssymb, unsortedadress]{revtex4-1}

 \usepackage{booktabs}
\usepackage{graphicx}
\usepackage{epstopdf}
\usepackage{dcolumn}
\usepackage{float}
\usepackage[normalem]{ulem}
\usepackage{color}
\usepackage{dcolumn}
 \usepackage{multirow}
\usepackage{subfigure}
\usepackage{cancel}
\newcommand{\be}{\begin{equation}}

\newcommand{\ee}{\end{equation}}
\newcommand{\bea}{\begin{eqnarray}}

\newcommand{\eea}{\end{eqnarray}}

\newcommand{\pup}{p^\uparrow}

\newcommand{\bfp}{\mbox{\boldmath $p$}}

\newcommand{\that}{\hat{t}} 
\newcommand{\uhat}{\hat{u}}
\newcommand{\shat}{\hat{s}}

\newcommand{\bkp}{\mbox{$\mathbf{k}_\perp$}}
\def\lsim{\mathrel{\rlap{\lower4pt\hbox{\hskip1pt$\sim$}}\raise1pt\hbox{$<$}}}
\def\gsim{\mathrel{\rlap{\lower4pt\hbox{\hskip1pt$\sim$}}\raise1pt\hbox{$>$}}}

\begin{document}

\title{Transverse single-spin asymmetry in the low-virtuality leptoproduction of open charm as a probe of the gluon Sivers function\\}

\author{Rohini M. Godbole}
\email{rohini@chep.iisc.ernet.in}
\author{Abhiram Kaushik}
\email{abhiramb@chep.iisc.ernet.in}
\affiliation{Centre for High Energy Physics, Indian Institute of Science, Bangalore, India.}

\date{\today}

\author{Anuradha Misra}
\email{misra@physics.mu.ac.in}

\affiliation{Department of Physics, University of Mumbai, Mumbai, India.}

\begin{abstract}

We study the low-virtuality inclusive leptoproduction of open charm, $p^\uparrow l\rightarrow D^0+X$ as a probe of the gluon Sivers function. We perform the analysis in a generalised parton model framework. At leading order, this process is sensitive only to the gluon content of the proton. Hence any detection of a transverse single-spin asymmetry in this process would be clear indication of a non-zero gluon Sivers function (GSF). Considering COMPASS and a future Electron-Ion Collider (EIC), we present predictions for asymmetry using fits for the GSF available in literature. Predictions for peak asymmetry values lie in the range of 0.8\% to 13\%. We also present estimates of the upper bound on the asymmetry as obtained with a maximal gluon Sivers function. Further, for the case of the Electron-Ion Collider, we evaluate the asymmetry in the muons decaying from the $D$-meson and find that the asymmetry is well preserved in the kinematics of the muons. Peak values of the muon asymmetry are close to those obtained for the $D$-meson and lie in the range $0.75\%$ to 11\%.

\end{abstract}

\maketitle

\section{\label{intro}Introduction}
Transverse single-spin asymmmetries (SSA) can provide crucial information on the three-dimensional structure of hadrons and have hence been a subject of increasing interest in the past two decades. While such asymmetries have been observed since the mid-70s in the hadroproduction of pions, i.e., $p^\uparrow p\to\pi+X$ ~\cite{Dick:1975ty, Klem:1976ui, Dragoset:1978gg}, the past few years have provided a large amount of high quality data on SSAs in a wide variety of processes such as $p^\uparrow p\to\pi+X$, $p^\uparrow p\to K^\pm+X$, $p^\uparrow p\to J/\psi+X$, $lp^\uparrow\to\pi+X$, $lp^\uparrow\to K+X$ etc., (see Refs.~\cite{DAlesio:2007bjf, Barone:2010zz} for reviews on the subject). A theoretical approach based on factorisation in terms of a hard-part and transverse momentum dependent (TMD) parton distribution functions (PDFs) and fragmentation function (FFs) has been formally established for processes which have  two scales: a hard, high energy scale such as the virtuality of the photon in the Drell-Yan process and a relatively soft scale of the order of $\Lambda_\text{QCD}$, such as the transverse momentum of the Drell-Yan lepton-pair. Another approach based on factorisation in terms of twist-3 parton correlators has been shown to be valid for the description of SSAs in processes with a single hard scale such as the transverse momentum of a pion in hadronic collisions.

Despite the absence of a formal proof, a lot of work has been done on a TMD description of single hard-scale processes under the assumption of factorisation, in what is generally referred to as the generalised parton model (GPM) approach. This approach has been quite succesful in describing unpolarised cross-sections in the hadroproduction of pions~\cite{DAlesio:2004eso}. The leading-order (LO) GPM is able to describe (upto a K-factor) experimental data on pion production in high energy hadron-hadron collisions better than either the LO or the NLO collinear pQCD. It is also able to provide a good description of data on SSA in $p^\uparrow p\to \pi+X$ at widely different c.o.m energies~\cite{DAlesio:2007bjf}. One of the important transverse-momentum-dependent distributions which can lead to SSAs is the Sivers distribution~\cite{Sivers:1989cc, Sivers:1990fh}, which encodes the correlation between the azimuthal anisotropy in the distribution of an unpolarised parton and the spin of its parent hadron. This anisotropy in the parton's transverse momentum distribution can lead to an azimuthal anisotropy in the distribution of the inclusive final state, i.e., a SSA. Fits of the $u$ and $d$ quark Sivers functions (QSF) obtained using data on $A_N(p^\uparrow p\to \pi+X)$ at E704 ($\sqrt{s}=19.4$ GeV), do a good job of describing the main features of the asymmetry observed at STAR ($\sqrt{s}=200$ GeV)~\cite{DAlesio:2007bjf}. While the quark Sivers functions have been widely studied over the years, the gluon Sivers function (GSF) still remains poorly understood.

A first indirect estimate of the gluon Sivers function in a GPM framework was obtained in Ref.~\cite{DAlesio:2015fwo}. The analysis consisted of fitting the GSF to midrapidity data on SSA in pion production at RHIC. In this analysis, the quark contribution to the SSA was calculated using quark Sivers functions (QSF) as extracted from semi-inclusive deep inelastic scattering data. The GSF fits obtained by this analysis predict asymmetries much smaller than allowed by the positivity bound on the GSF, viz. twice the unpolarised TMD gluon distribution. On the other hand, a recent study of large-$p_T$ hadron pair production in COMPASS indicates a substantial negative gluon Sivers asymmetry for both proton and deuteron targets~\cite{Adolph:2017pgv}. Since large-$p_T$ hadron pairs are produced through the photon-gluon fusion, this process is indeed sensitive to the GSF. However the final state also receives contributions from the QCD Compton process, which is quark initiated.  Hence an extraction of the GSF using large-$p_T$ hadron production is contaminated by the quark contributions to the SSA and would therefore depend on the extent to which these different processes can be separated in a data sample. With this being the first significant evidence for a non-zero GSF, it is important to study processes such as closed and open charm production which probe the gluon channel cleanly and directly.  

In this work, using a GPM approach, we study the low-virtuality leptoproduction ($Q^2\approx0$) of open-charm as a possible probe of the poorly understood gluon Sivers function (GSF). At the leading-order (LO) of this process, the production of open-charm happens only via photon-gluon fusion (PGF), making it a direct probe of the gluon content of the proton. The GPM study of open-charm production as a probe of the gluon Sivers function was first proposed in Ref.~\cite{Anselmino:2004nk} for the process $p^\uparrow p\to D^0+X$. In that study they considered two extreme scenarios for the GSF: zero and maximal. The term `maximal' here refers to the Sivers function with its positivity bound of twice the unpolarised TMD ($|\Delta^Nf_{i/p^\uparrow}(x,\mathbf{k}_\perp)|/2f_{i/p}(x,\mathbf{k}_\perp)\leq1$), saturated for all values of $x$ --- we shall refer to this as the `saturated' Sivers function henceforth. Their study indicated that a measurement of SSA at RHIC for this process can give a direct indication of a nonzero gluon Sivers function. Further in Ref.~\cite{Godbole:2016tvq} we caculated the SSA for the same process (open charm hadroproduction) using the fits of Ref.~\cite{DAlesio:2015fwo} and found that these fits predict sizeable, measurable asymmetries.

The low-virtuality leptoproduction of $J/\psi$ has also been suggested as a probe of the GSF~\cite{Godbole:2012bx,Godbole:2013bca, Godbole:2014tha}. However, leptoproduction of open-charm may have some more advantages over the above mentioned processes. Firstly, unlike the case with $p^\uparrow p\to D^0+X$, one need not worry about possible factorisation breaking initial state interactions. $p^\uparrow l\to D^0+X$ would have the same initial/final state interactions as SIDIS, for which TMD factorisation has been established. A study of SSA in this process might therefore complement studies of SSA in SIDIS and $lp^\uparrow\to h+X$~\cite{DAlesio:2017nrd, DAlesio:2017jdm} by providing an additional handle on the gluon Sivers function. Secondly, open-charm production is free from dependence on production model, as is the case with closed-charm~\cite{Yuan:2008vn}.


We therefore consider the process $lp^\uparrow\to D^0+X$ in the low-virtuality regime in a GPM framework and see how it could serve as a probe of the GSF at both the COMPASS experiment and a future Electron-Ion Collider (EIC). While present data on open-charm production in COMPASS is limited due to statistics, the proposed Electron-Ion Collider~\cite{Accardi:2012qut} would have a significantly higher luminosity and should be able to provide better data on open/closed charm production. For both experiments, we present estimates for the maximum magnitude of SSA as obtained using the saturated GSF, and also the expected values of SSA obtained using the fits of Ref.~\cite{DAlesio:2015fwo}.

In section II, we give the expressions for the relevant quantities in the GPM framework. In section III, we give parametrisations for the different TMDs used, and in section IV, we discuss results for both the COMPASS and the EIC kinematics.

\section{The GPM Formalism}
In this work, we are concerned with the single-spin asymmetry in the low-virtuality leptoproduction of open charm, 
\be
A_N=\frac{d\sigma^\uparrow-d\sigma^\downarrow}{d\sigma^\uparrow+d\sigma^\downarrow}
\label{SSA}
\ee
where $d\sigma^{\uparrow( \downarrow)}$ is the invariant differential cross-section for the process $p^{\uparrow(\downarrow)} l\to D+X$ with the spin of the transversely polarised proton being aligned in the $\uparrow$($\downarrow$) direction with respect to the production plane. Here, $\uparrow$ would be the $+y$ direction in a frame where the polarised proton is moving along the $+z$ direction and the meson is produced in the $xz$ plane. Note that this is the convention for collider experiments (such as EIC). In case of fixed target experiments (such as COMPASS), by convention, the polarised proton would be considered to be moving along the $-z$ direction with everything else remaining the same (cf.~Fig.~\ref{conventions}).

Following the treatment of open-charm hadroproduction~\cite{Anselmino:2004nk}, we can write the denominator and numerator of Eq.~1 as,
\bea
d\sigma ^\uparrow + d\sigma ^\downarrow &=& 
\frac{E_D \, d\sigma^{p^\uparrow l \to DX}} {d^{3} \bfp_D} +
\frac{E_D \, d\sigma^{p^\downarrow l \to DX}} {d^{3} \bfp_D}
= 2 \, \frac{E_D \, d\sigma^{pl \to DX}} {d^{3} \bfp_D} 
\label{denominator}\\ 
&& \hspace*{-3.0cm} = \>
2\int dx_g \, dx_\gamma  \, dz \, d^2 \mathbf{k}_{\perp g} \, d^2 \mathbf{k}_{\perp \gamma} \, 
d^3 \mathbf{k}_{D} \, 
\delta (\mathbf{k}_{D} \cdot \hat{\bfp}_c) \, 
\delta (\hat s +\hat t +\hat u - 2m_c^2) \> 
{\mathcal C}(x_g,x_\gamma,z,\mathbf{k}_D) \nonumber \\
&& \hspace*{-3.0cm} \times ~
f_{g/p}(x_g,\mathbf{k}_{\perp g}) \>  f_{\gamma/l}(x_\gamma, 
\mathbf{k}_{\perp \gamma}) \>
\frac{d \hat{\sigma}^{g\gamma \to c \bar c}}
{d\hat t}(x_g, x_\gamma, \mathbf{k}_{\perp g}, \mathbf{k}_{\perp \gamma}, \mathbf{k}_D) \>
 D_{D/c}(z,\mathbf{k}_D) \> \nonumber 
\eea
and
\bea
d\sigma ^\uparrow - d\sigma ^\downarrow &=& 
\frac{E_D \, d\sigma^{p^\uparrow l  \to DX}} {d^{3} \bfp_D} -
\frac{E_D \, d\sigma^{p^\downarrow l\to DX}} {d^{3} \bfp_D}  
\label{numerator}\\
&& \hspace*{-3.0cm} = \>
\int dx_g \, dx_\gamma  \, dz \, d^2 \mathbf{k}_{\perp g} \, d^2 \mathbf{k}_{\perp \gamma} \, 
d^3 \mathbf{k}_{D} \, 
\delta (\mathbf{k}_{D} \cdot \hat{\bfp}_c) \, 
\delta (\hat s +\hat t +\hat u - 2m_c^2) \> 
{\mathcal C}(x_g,x_\gamma,z,\mathbf{k}_D) \nonumber \\
&& \hspace*{-3.0cm} \times~
 \Delta ^N f_{g/\pup}(x_g,\mathbf{k}_{\perp g}) \>  f_{\gamma/l}(x_\gamma, 
\mathbf{k}_{\perp \gamma}) \>
\frac{d \hat{\sigma}^{g\gamma \to c \bar c}}
{d\hat t}(x_g, x_\gamma, \mathbf{k}_{\perp g}, \mathbf{k}_{\perp \gamma}, \mathbf{k}_D) \>
 D_{D/c}(z,\mathbf{k}_D). \> \nonumber 
\eea
In the above expressions, $x_{g(\gamma)}$ is the light-cone momentum fraction of the incoming gluon (photon) with the $z$-axis along the parent proton (lepton) direction, $z=p_D^+/p_c^+$ is the light-cone momentum fraction of the $D$-meson with the $z$-axis along the fragmenting charm quark direction, $\mathbf{k}_{g(\gamma)}$ is the intrinsic transverse momentum of the gluon (photon) with respect to the parent particle direction, $\mathbf{k}_D$ is the transverse momentum with which the meson fragments from the charm quark, $\hat\bfp$ is the unit vector along the heavy quark direction, $m_c$ is the charm quark mass, and $\shat$, $\that$ and $\uhat$ are the Mandelstam variables for the photon-gluon fusion process $\gamma g\to c\bar{c}$.

$\Delta^Nf_{g/p^\uparrow}(x,\mathbf{k}_\perp)$ and $f_{g/p}(x,\mathbf{k}_\perp)$ stand for the gluon Sivers function and unpolarised TMD respectively. $f_{\gamma/l}(x,\bkp)$ is the transverse-momentum-dependent distribution of quasi-real photons in an unpolarised lepton, and $D_{D/c}(z,\mathbf{k}_D)$ is the transverse-momentum-dependent fragmentation function. We will discuss the functional forms for all these distributions in Sec. III.

As mentioned earlier, the Sivers function, $\Delta^Nf_{i/p^\uparrow}(x,k_\perp;Q)$ describes the azimuthal anisotropy in the transverse momentum distribution of an unpolarised parton in transversely polarised hadron, and we have
\bea
f_{i/h^\uparrow}(x,\mathbf{k}_\perp,\mathbf{S};Q)&=&f_{i/h}(x,k_\perp;Q)+\frac{1}{2}\Delta^N f_{i/h^\uparrow}(x,k_\perp;Q)\frac{\epsilon_{ab}k_\perp^a S^b}{k_\perp}\nonumber \\
&=&f_{i/h}(x,k_\perp;Q)+\frac{1}{2}\Delta^N f_{i/h^\uparrow}(x,k_\perp;Q)\cos\phi_{\perp}
\label{Sivers}
\eea
where $\mathbf{k}_\perp=k_\perp(\cos\phi_\perp, \sin\phi_\perp)$. In a generalised parton model (GPM) description of this process, the only possible source of an asymmetry would be a non-zero gluon Sivers function. Since photon-gluon fusion results in unpolarised final state quarks, there cannot be any contribution from the Collins effect, which would require transversely polarised final state quarks.

The partonic cross-section for photon-gluon fusion into a heavy quark pair is given by~\cite{Babcock:1977fi},
\bea
\frac{d\hat{\sigma}^{g\gamma \to c \bar c}}{d\hat t}&=&\frac{4\pi}{8\times9}~\frac{\alpha_\text{em}\alpha_s}{ \shat^{2}(\that-m_c^2)^{2}(\uhat-m_c^2)^{2}}[-(\that-\uhat)^4-4\shat(\that+\uhat)(\that-\uhat)^2
\\ \nonumber &-&4\shat^2\left((\that-\uhat)^2+2(\that+\uhat)^2\right)-12\shat^3(\that+\uhat)-3\shat^4]
\eea
where the Mandelstam variables are defined in the usual way,
\be
\hat s = (P_g+P_\gamma)^2;\> \hat t = (P_g-P_c)^2;\>\hat u = (P_\gamma-P_c)^2.
\ee
The factor ${\mathcal C}(x_g,x_\gamma,z,\mathbf{k}_D)$ in Eqs.~2 and 3 contains the parton flux and the Jacobian relating the partonic phase-space to the mesonic phase-space. It is give by,
\be
{\mathcal C}(x_g,x_\gamma,z,\mathbf{k}_D)=\frac{\hat s}{\pi z^2}\,\frac{\hat s}{x_g x_\gamma s}\,
\frac{ \left( E_D+\sqrt{\mathbf{p}_D^2 - \mathbf{k}_{\perp D}^2} \right)^2}
{4(\mathbf{p}_D^2 - \mathbf{k}_{\perp D}^2)} \,
\left[1- \frac{z^2 m_c^2}
{ \left( E_D+\sqrt{\mathbf{p}_D^2 - \mathbf{k}_{\perp D}^2} \right)^2}\right]^2
\ee
The on-shell condition $\shat+\that+\uhat=2m_c^2$ in Eqs.~2 and 3, gives a quartic equation in $z$. $z$ can then be fixed by using this equation as shown in  Ref.~\cite{Godbole:2016tvq}.

The delta function $\delta (\mathbf{k}_{D} \cdot \hat{\bfp}_c)$ in Eqs.~2 and 3 ensures that the region of integration for $\mathbf{k}_D$ is confined to the two-dimensional plane perpendicular to the direction of the charm quark i.e., 
\be
\int d^3\mathbf{k}_D~\delta (\mathbf{k}_{D} \cdot \hat{\bfp}_c)D_{D/c}(z,\mathbf{k}_D)...=\int d^2\mathbf{k}_{\perp D}D_{D/c}(z,\mathbf{k}_{\perp D})...
\ee
where $\mathbf{k}_{\perp D}$ represents values of transverse momenta on the allowed plane.

An outline of the treatment of the parton level kinematics and the TMD fragmentation is given in Appendix A.

\section{Parametrisation of the TMDs}
Since we give predictions using the GSF fits of Ref.~\cite{DAlesio:2015fwo}, for consistency we have to use the unpolarised gluon TMD and Sivers function used therein. We use standard factorised Gaussian form for the unpolarised gluon TMD,
\be
f_{g/p}(x,k_\perp;Q)=f_{g/p}(x,Q)\frac{1}{\pi\langle k_\perp^2\rangle}e^{-k_\perp^2/\langle k_\perp^2\rangle}
\ee
with $\langle k_\perp^2\rangle=0.25\text{ GeV}^2$. 

For the photon distribution $f_{\gamma/l}(x,\mathbf{k}_\perp)$, we consider two cases:
\begin{enumerate}
\item The Weizsacker-Williams distribution of quasi-real photons with a Gaussian transverse momentum spread~\cite{Godbole:2012bx,Godbole:2013bca, Godbole:2014tha},
\be
f_{\gamma/l}(x_\gamma,\mathbf{k}_{\perp \gamma};s)=f_{\gamma/l}(x_\gamma,s)\frac{1}{\pi\langle k_{\perp\gamma}^2\rangle}e^{-k_{\perp g}^2/\langle k_{\perp\gamma}^2\rangle}\hspace*{0.5cm}\text{(Gaussian WW)}
\ee
where the Weizsacker-Williams distribution is given by~\cite{Brodsky:1971ud, Terazawa:1973tb,Kniehl:1990iv},
\be
f_{\gamma/l}(x_\gamma,s)=\frac{\alpha_\text{em}}{\pi}\left(
\frac{1+(1-x_\gamma)^2}{x_\gamma}\left[\log\frac{\sqrt{s}}{2m_l}-\frac{1}{2}\right]
\right)
\ee
and the width of the Gaussian is $\langle k_{\perp\gamma}^2\rangle=0.1$ GeV$^2$.
\item The leading order result for the TMD distribution of photons in a lepton from Ref.~\cite{Bacchetta:2015qka},
\be
f_{\gamma/l}(x_\gamma,\mathbf{k}_{\perp \gamma})=\frac{\alpha_\text{em}}{2\pi^2}\frac{\mathbf{k}^2_{\perp\gamma}\left[1+(1-x_\gamma)^2\right]+m^2x_\gamma^4}{x_\gamma\left[\mathbf{k}^2_{\perp\gamma}+m^2x_\gamma^2\right]^2}\hspace*{0.5cm}\text{(Photon TMD)}
\ee
where $m$ is the mass of the lepton.
\end{enumerate}
The first choice, which we will refer to as Gaussian WW, was used in earlier studies of low-virtuality leptoproduction by us~ \cite{Godbole:2012bx,Godbole:2013bca, Godbole:2014tha} (and also in an analysis of low-$Q^2$ contributions to $ep^\uparrow\to h+X$~\cite{DAlesio:2017nrd}, but without the Gaussian spread) when first-principles result for the photon TMD distribution was not available. The second choice, which we will refer to as Photon TMD, is the first analytical result available in literature for the transverse-momentum-dependent distribution of photons in a lepton~\cite{Bacchetta:2015qka}. Here we present results using both choices for completeness.

As with the unpolarised densities, we take the transverse-momentum-dependence of the FF to be Gaussian,
\be
D_{D/c}(z,\mathbf{k}_{D})=D_{D/c}(z)\frac{1}{\pi\langle k^2_{\perp D}\rangle}e^{-k_D^2/\langle k^2_{\perp D}\rangle}
\ee
with $\langle k^2_{\perp D}\rangle=0.25$ GeV$^2$.

The gluon Sivers function is parametrised as follows~\cite{DAlesio:2015fwo},
\be
\Delta ^N f_{g/\pup}(x,k_{\perp};Q)=2\mathcal{N}_g(x)f_{g/p}(x,Q)\frac{\sqrt{2e}}{\pi}\sqrt{\frac{1-\rho}{\rho}}k_\perp \frac{e^{-k^2_\perp/\rho\langle k^2_\perp\rangle}}{\langle k^2_\perp\rangle^{3/2}}
\ee
with $0<\rho<1$. $\mathcal{N}_g(x)$ parametrises the $x$-dependence of the GSF and is generally written as
\be
\mathcal{N}_g(x)=N_gx^{\alpha_g}(1-x)^{\beta_g}\frac{(\alpha_g+\beta_g)^{\alpha_g+\beta_g}}{\alpha_g^{\alpha_g}\beta_g^{\beta_g}}
\ee
It must obey $|\mathcal{N}_g(x)|<1$ in order for the Sivers function to satisfy the positivity bound,
\be
\frac{|\Delta^Nf_{g/p^\uparrow}(x,\mathbf{k}_\perp)|}{2f_{g/p}(x,\mathbf{k}_\perp)}\leq 1\>\forall \>x, \mathbf{k}_\perp.
\ee

\begin{table}[t]
\centering
\begin{tabular}{|l|l|l|l|l|l|l|}
\hline
SIDIS1 & \multicolumn{2}{l|}{$N_g=0.65$} & $\alpha_g=2.8$ & $\beta_g=2.8$ & $\rho=0.687$ & \multirow{2}{*}{$\langle k^2_\perp\rangle=0.25$ GeV$^2$} \\ \cline{1-6}
SIDIS2 & \multicolumn{2}{l|}{$N_g=0.05$} & $\alpha_g=0.8$ & $\beta_g=1.4$ & $\rho=0.576$ &                                                        \\ \cline{1-7}
\end{tabular}
\caption{Parameters of the GSF fits from Ref.~\cite{DAlesio:2015fwo}.}
\label{SIDIS-gluon-fits}
\end{table}

In this work, for the predictions we consider two options for the gluon Sivers function: 
\begin{enumerate}
\item the Sivers function with the positivity bound saturated, i.e., $\mathcal{N}_g(x)=1$ and $\rho=2/3$.
\item the SIDIS1 and SIDIS2 extractions of the Sivers function from Ref.~\cite{DAlesio:2015fwo}.
\end{enumerate}

As mentioned in the introduction, we will refer to the first choice as the `saturated' Sivers function. It would give an upper bound on the asymmetry for a fixed width $\langle k^2_\perp\rangle$. The parameter $\rho$ is set to $2/3$ in order to maximize the first $k_\perp$-moment of the Sivers function, following Ref.~\cite{DAlesio:2010sag}. It must be kept in mind though, that this cannot be treated as giving an absolute upper bound on $A_N$ --- an increased width $\langle k^2_\perp\rangle$, for a fixed value of $\rho$, naturally would result in an increased asymmetry since the effects of the parton transverse momenta are more pronounced.

The SIDIS1 and SIDIS2 GSFs from Ref.~\cite{DAlesio:2015fwo} are the first (and so far, only) available extractions of the GSF in a GPM framework. They were obtained by fitting to PHENIX data on $A_N$ in inclusive pion production in the midrapidity region at RHIC. The QSFs used in these extractions, also labelled SIDIS1 and SIDIS2 respectively, were fit to data on semi-inclusive deep inelastic scattering. The SIDIS1 QSF set~\cite{Anselmino:2005ea} (which was used in the extraction of the SIDIS1 GSF) was fitted to data on pion production in HERMES and positive hadron production in COMPASS with fragmentation functions by Kretzer~\cite{Kretzer:2000yf}. It contains only the $u$ and $d$ quark Sivers functions since the data was not sensitive to sea quark contributions. The SIDIS2 QSF set~\cite{Anselmino:2008sga} was fitted to flavour segregated data on pion and kaon production from HERMES and COMPASS and hence included sea quark Sivers functions as well. It used fragmentation functions by de Florian, Sassot and Stratmann (DSS)~\cite{deFlorian:2007aj}.

Both QSF sets give a good description of their respective SIDIS data sets. Furthermore both the GSFs (taken along with their associated QSF sets) describe the data on $A_N$ in midrapidity pion production equally well. Despite this the two fits show very different $x$-dependencies, with SIDIS1 being larger in the moderate-$x$ region and SIDIS2 being larger in the low-$x$ region. The values of the parameters of the two GSF fits are given in Table I.

\section{Results}

\begin{figure}[h]
\begin{center}
\includegraphics[width=0.45\linewidth]{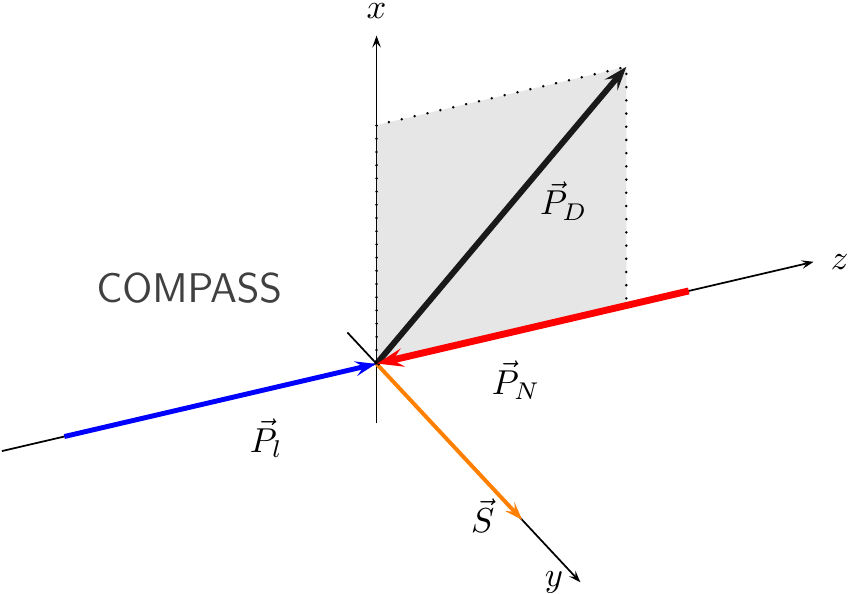}
\includegraphics[width=0.45\linewidth]{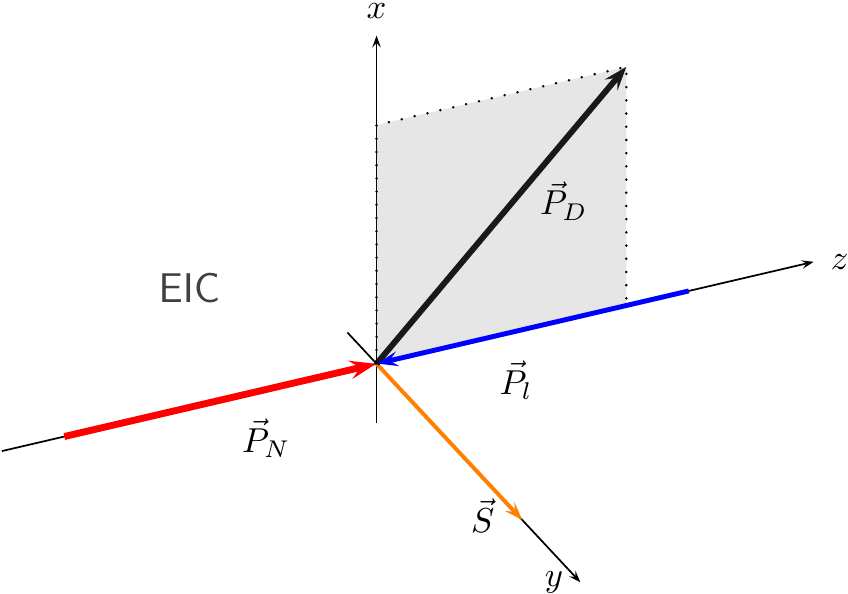}
\caption{Kinematics for COMPASS (left) and EIC (right). $\vec{P}_N$ is the proton momentum  and $\vec{S}$ is its spin orientation. $\vec{P}_l$ is the lepton momentum. The $D$-meson momentum, $\vec{P}_D$ is taken to be on the $x\text{--}z$ plane.}
\label{conventions}
\end{center}
\end{figure}

In this section we present results on the unpolarised cross-section and SSA for COMPASS and EIC kinematics. Before going into the  results, we should first make a note on the differing kinematic conventions of the two experiments: As COMPASS is a fixed target experiment, by convention the lepton is taken to be along the $+z$ direction. This means that, in the definition of $A_N$ in Eq.~1, keeping the conventions for proton spin direction and production plane the same, positive $x_F$ and $\eta$ correspond to the backward hemisphere of the proton, whereas negative $x_F$ and $\eta$ correspond to the forward hemisphere of the proton. Note that this convention differs from that adopted Sec.~II where, following the RHIC convention, the {\it proton} is taken to be moving along the $+z$ direction. Since EIC, like RHIC, is also a collider experiment, we shall use the same convention for it. In the interest of clarity, the conventions used for the two experiments are illustrated in Fig.~\ref{conventions}.

Please note that since we are interested in quasi-real photoproduction, we have put a cut, $Q^2<1$ GeV$^2$, where $Q^2=-(P_l-P_{l'})^2$ is the photon virtuality. This was motivated by the COMPASS antitagging cuts.  In regions of large photon transverse momenta, the lepton-photon vertex becomes $hard$ and the photon becomes off-shell. Hence one cannot use hard-parts defined for on-shell initial and final states.  The cut on the photon virtuality $Q^2$ can be implemented by considering its relation to $k_{\perp\gamma}$ and $x_\gamma$, $Q^2=k_{\perp\gamma}^2\left(1+\frac{x_\gamma}{1-x_\gamma}\right)$. In this work, all results associated with both COMPASS as well as EIC were obtained with the $Q^2<1$ GeV$^2$ cut. When using the Gaussian WW approximation, this cut does not make a huge difference since the steeply falling  $k_{\perp\gamma}$-dependence for $k_{\perp\gamma}>\sqrt{\langle k_{\perp\gamma}^2\rangle}$ prevents large contributions from regions of large virtuality. However, this is not the case with the Photon TMD as it has a much longer tail due to its $1/k_{\perp\gamma}^2$ dependence. One must also note here that the opposite is true in the low-$k_{\perp\gamma}$ region. At low $k_{\perp\gamma}$ the Photon TMD falls off very sharply with increasing ${k}_{\perp\gamma}$ whereas the Gaussian WW approximation, by virtue of being a Gaussian, has a  flat $k_T$-dependence at very low $k_T$.

The numerical results were obtained using the GRV98LO set for the collinear gluon density and for the collinear part of the FF, the LO parametrisation of the $c\to D^0$ fragmentation function by Kniehl and Kramer~\cite{Kniehl:2006mw} was used. The QCD scale was chosen to be $Q^2=m_D^2+P_T^2$.

\subsection{COMPASS}
The COMPASS experiment is a fixed target experiment involving a 160 GeV muon beam colliding on a proton target with a centre of mass energy $\sqrt{s}=17.4$ GeV. The COMPASS spectrometer covers hadrons in the $l\text{-}p$ c.o.m frame pseudorapidity range $-0.1<\eta_h<2.4$ and detects $D^0$ mesons through their $D^0\to K^-\pi^+$ decay mode. The geometry of the detector allows a proper reconstruction of the $D^0$'s produced only in the backward hemisphere of the proton and hence we restrict our analysis to the $x_F, ~\eta>0$ region.

\begin{figure}[t]
\begin{center}
\vspace*{-2cm}
\includegraphics[width=\linewidth]{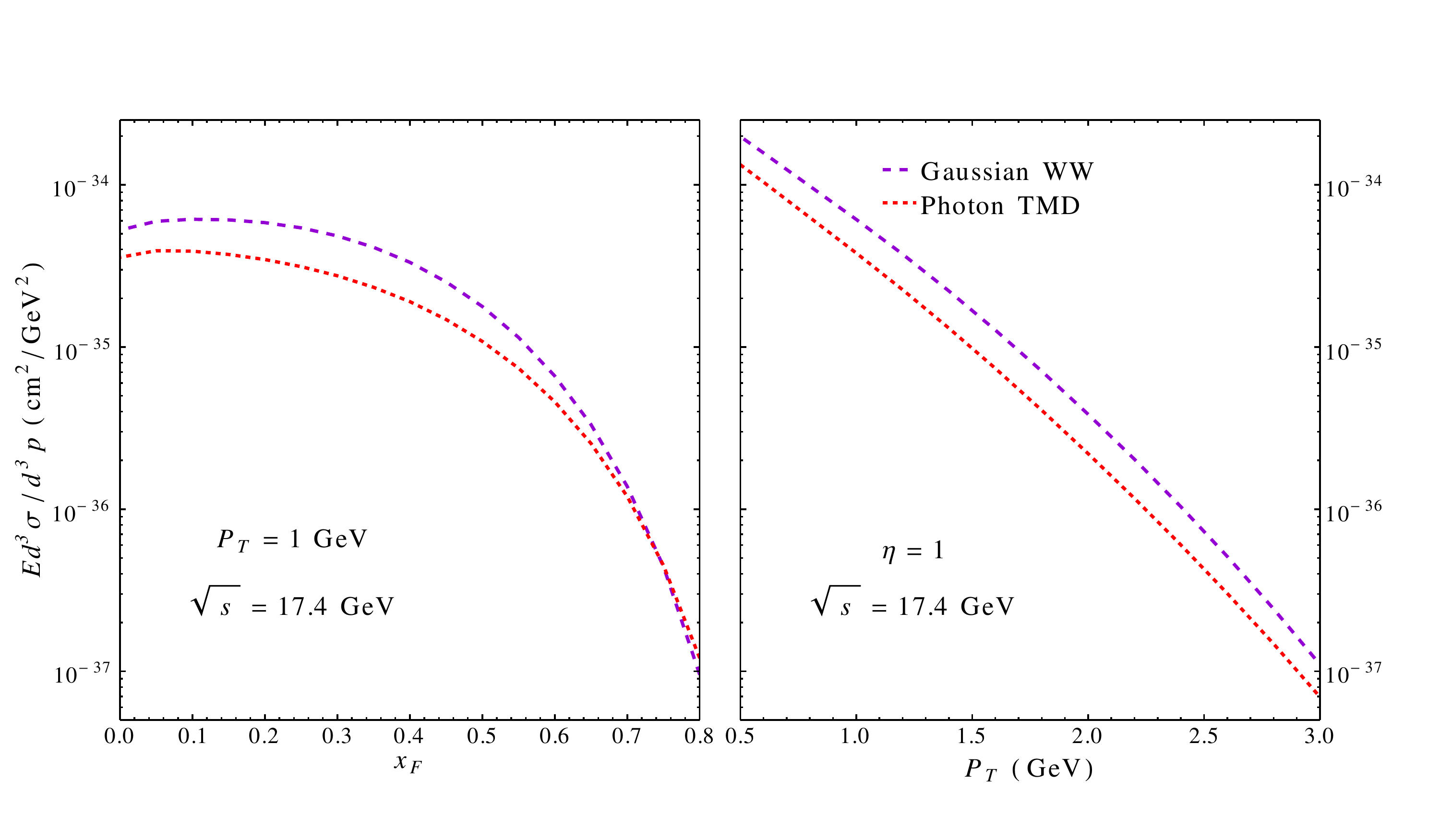}
\vspace*{-2cm}
\caption{Unpolarized cross-section at COMPASS as a function of $x_F$ (at fixed $P_T$, left panel) and $P_T$ (at fixed $\eta$, right panel).}
\label{COMPlogcs}
\end{center}
\end{figure}

In Fig.~\ref{COMPlogcs}, we show results for the unpolarised invariant cross-section using both the Gaussian WW approximation and the Photon TMD with the $Q^2<1$ GeV$^2$ cut. We show the cross-section as a function of $x_F$ at fixed $P_T=1$ GeV (left panel) and as a function of $P_T$ at a fixed pseudorapidity $\eta=1$ (right panel). At a fixed $P_T$, the cross-sections obtained with both the Gaussian WW and Photon TMD vary by two orders of magnitude in the region $0<x_F<0.8$. The Photon TMD result is generally smaller than the Gaussian WW result by 30-40\%, except at very large $x_F$ where both become comparable. At fixed pseudorapidity, for both choices of the photon density, the cross-section decreases by three orders of magnitude with increasing $P_T$ in the range $0.5<P_T<3.0$ GeV. The cross-section obtained with the Photon TMD is smaller by roughly 30-40\% over the entinre $P_T$ range.  For a larger value of the width of unpolarised gluon TMD, viz. $\langle k_\perp^2\rangle=1$ GeV$^2$ instead of 0.25 GeV$^2$, the cross-section at fixed $P_T$ is not affected much whereas, the cross-section at fixed pseudorapidity spreads out in $P_T$ somewhat --- becoming smaller by 6\% at $P_T=0.5$ GeV and larger by 40\% at $P_T=3$ GeV --- as one would expect. Overall, cross-section estimates for COMPASS are not very sensitive to the unpolarised gluon TMD width. Varying the width of the TMD FF in the range $0<\langle k^2_{\perp D}\rangle<0.25$ GeV$^2$ also does not have any significant effect on the cross-section.

\begin{figure}[t]
\begin{center}
\vspace*{-2cm}
\includegraphics[width=\linewidth]{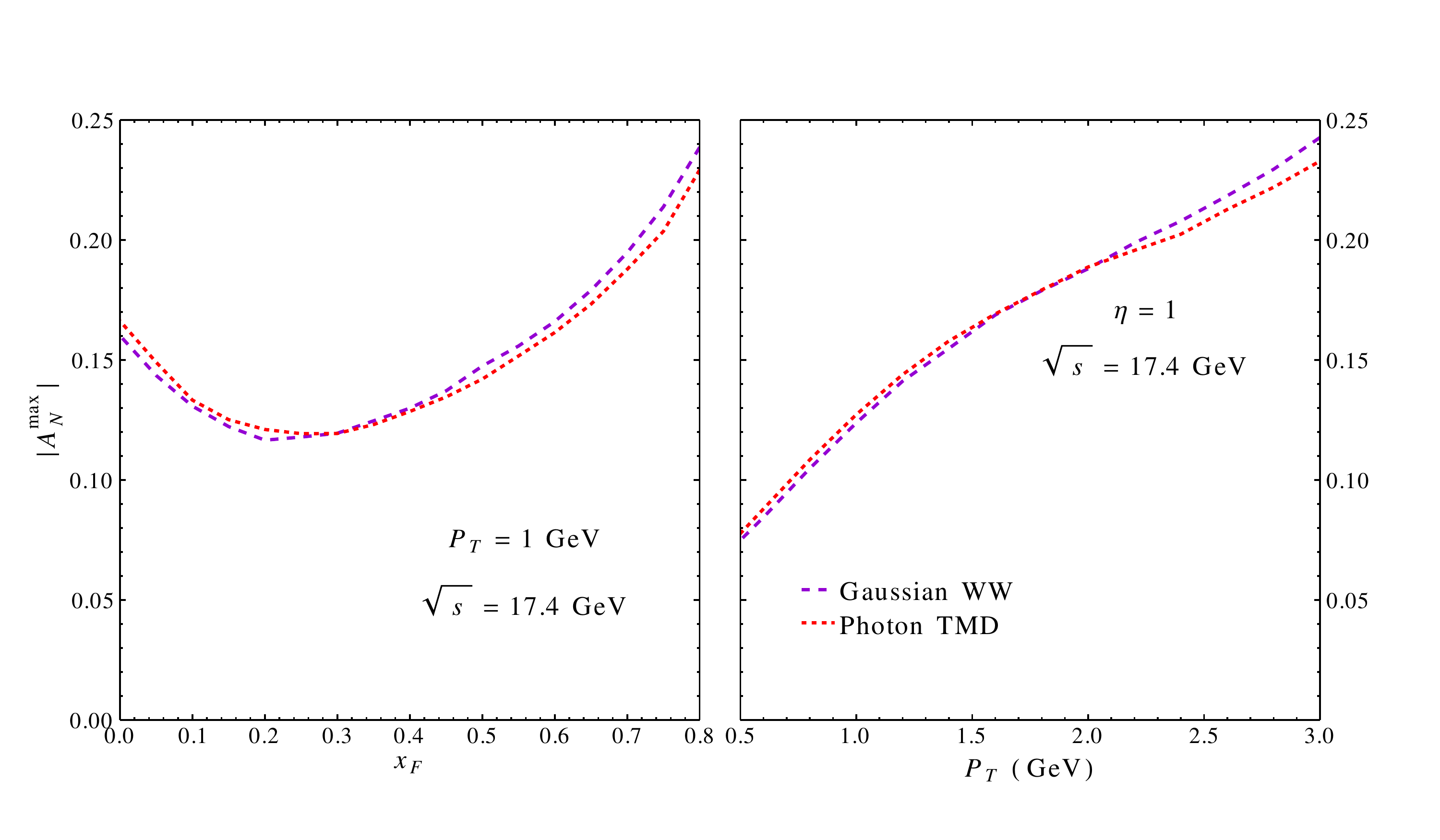}
\vspace*{-2cm}
\caption{SSA with saturated GSF at COMPASS as a function of $x_F$ (at fixed $P_T$, left panel) and $P_T$ (at fixed $\eta$, right panel).}
\label{COMPsatAN}
\end{center}
\end{figure}

Fig.~\ref{COMPsatAN} shows estimates for the maximal value of the magnitude of the asymmetry $|A^\text{max}_N|$, obtained by using the saturated gluon Sivers function viz., $\mathcal{N}_g(x)=1$, $\rho=2/3$. The results are presented  as a function of $x_F$ at fixed $P_T=1$ GeV (left panel) and as a function of $P_T$ at a fixed pseudorapidity $\eta=1$ (right panel). At fixed $P_T$, estimates of $|A^\text{max}_N|$  range from a minimum of about 12\% at $x_F\approx0.2-0.3$ to upto 24\% at $x_F=0.8$. At fixed  $\eta$, $|A^\text{max}_N|$ shows a general increase with the meson transverse momentum, ranging from around $8\%$ at $P_T=0.5$ GeV to $24\%$ at $P_T=3$ GeV. Both the Gaussian WW distribution and the Photon TMD give similar results with the former being slightly smaller at low $x_F$/$P_T$ and vice versa.

\begin{figure}[t]
\begin{center}
\vspace*{-1cm}
\includegraphics[width=\linewidth]{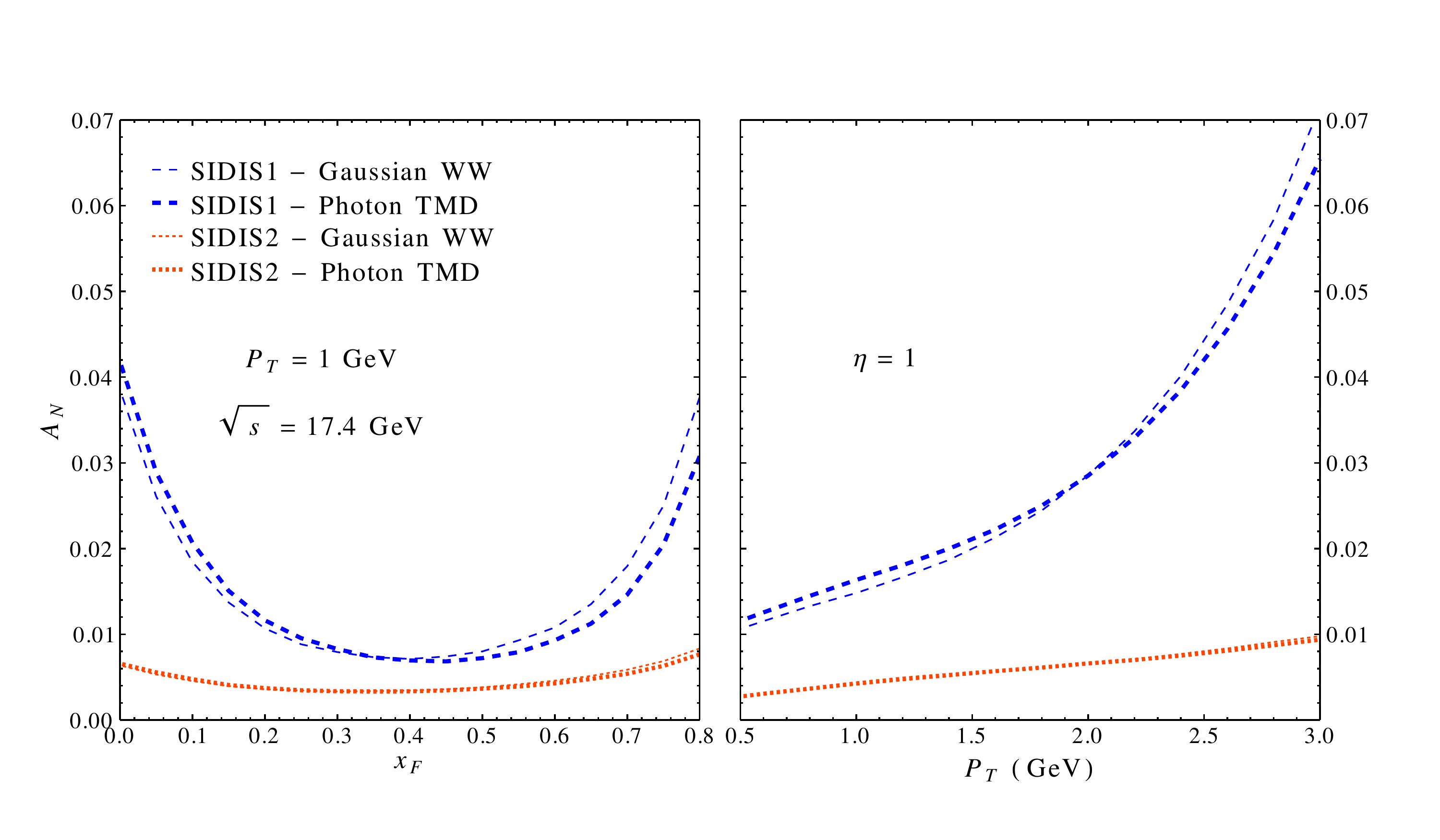}
\vspace*{-2cm}
\caption{SSA from GSF fits of Ref.~\cite{DAlesio:2015fwo} at COMPASS as a function of $x_F$ (at fixed $P_T$, left panel) and $P_T$ (at fixed $\eta$, right panel).}
\label{COMPsidisAN}
\end{center}
\end{figure}

Fig.~\ref{COMPsidisAN} shows the asymmetries obtained using the SIDIS1 and SIDIS2 fits~\cite{DAlesio:2015fwo}. As was the case with the saturated asymmetry, the results obtained with the Gaussian WW approximation and Photon TMD are generally similar, with the former being slightly smaller at low $x_F$/$P_T$ and vice versa. Both fits give asymmetry predictions much smaller than allowed by the positivity bound with SIDIS2 giving significantly smaller asymmetries than SIDIS1. This is because the kinematic regions we consider probe the region $0.08<x_g<0.5$, where SIDIS2 is much smaller than SIDIS1, as can be seen from the numbers in Table~\ref{SIDIS-gluon-fits}. At fixed $P_T$, SIDIS1 gives a peak asymmetry of $4.2\%$ at $x_F=0$ and SIDIS2 gives a peak asymmetry of $0.8\%$ at $x_F=0.8$. At fixed $\eta=1$, SIDIS1 gives a peak asymmetry of $7\%$ and SIDIS2 gives a peak asymmetry of almost $1\%$, both at $P_T=3.0$ GeV. We have verified that changes in the width of the TMD FF in the range $0<\langle k^2_{\perp D}\rangle<0.25$ GeV$^2$ do not alter the results for either SIDIS1 or SIDIS2 substantially and the general features of the $A_N$ predictions stay the same.

\subsection{EIC}

\begin{figure}[t]
\begin{center}
\vspace*{-2cm}
\includegraphics[width=\linewidth]{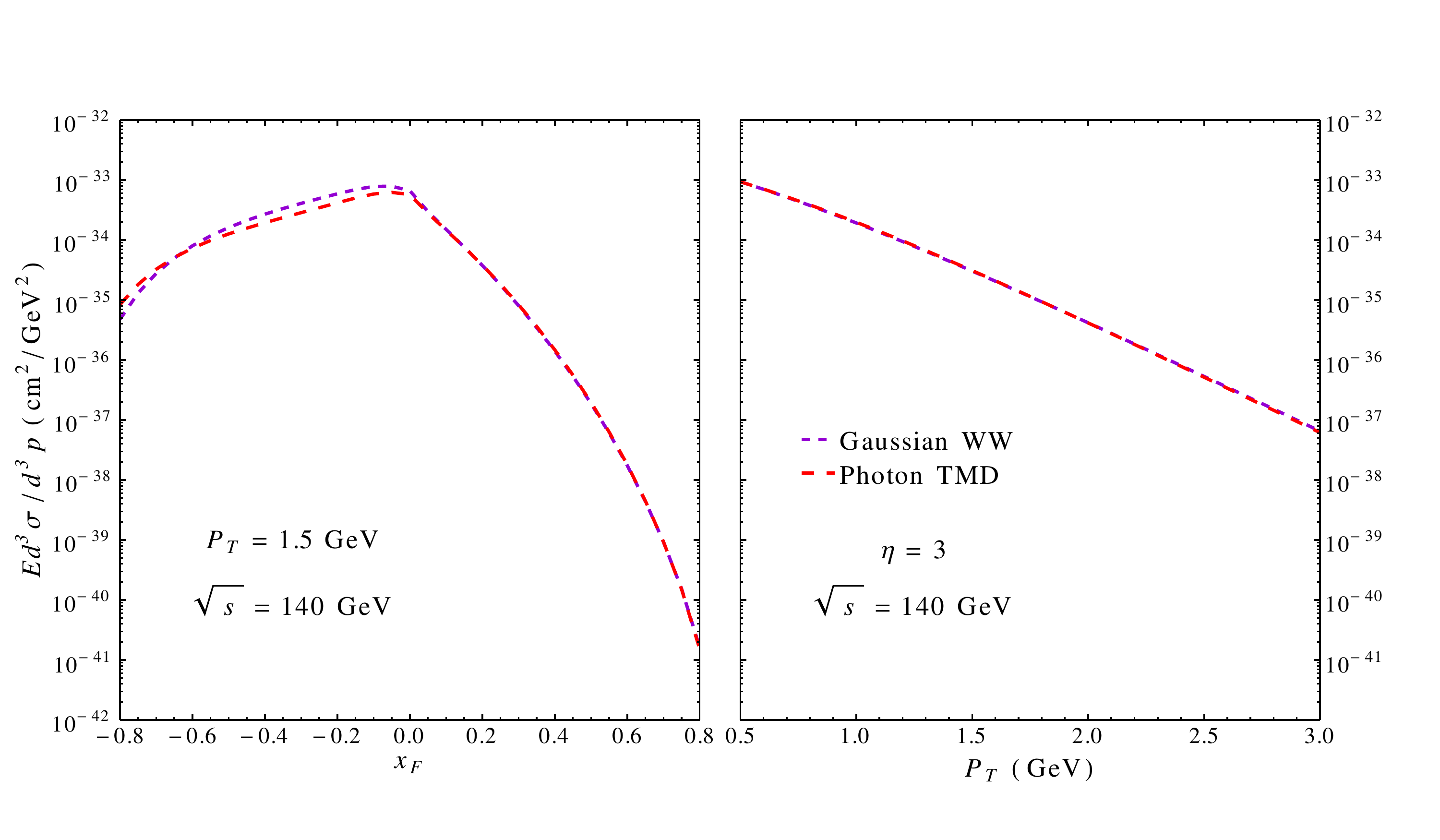}
\vspace*{-2cm}
\caption{Unpolarized cross-section at EIC as a function of $x_F$ (at fixed $P_T$, left panel) and $P_T$ (at fixed $\eta$, right panel).}
\label{EIClogcs}
\end{center}
\end{figure}

The Electron-Ion Collider (EIC) is a proposed experiment with colliding electron and proton/ion beams, with the possibility of both being polarised. It is meant to be capable of attaining high luminosities, with a centre of mass energy of upto 140 GeV in the $ep$ configuration.

In Fig.~\ref{EIClogcs}, we show results for the unpolarised invariant cross-section using both the Gaussian WW approximation and the Photon TMD with the same $Q^2<1$ GeV$^2$ cut as used for COMPASS. We show the cross-section as a function of $x_F$ at fixed $P_T=1.5$ GeV (left panel) and as a function of $P_T$ at a fixed pseudorapidity $\eta=3$ (right panel). At fixed $P_T$, in the forward region, the cross-section decreases with increasing $x_F$ by more than six orders of magnitude in the range $0<x_F<0.7$. In contrast, in the backward region, the decrease in the cross-section with increasing $|x_F|$, is only around one order of magnitude. This is because, for $x_F<0$ the gluon density is being probed in the small-$x$ region and both the Weiszacker-Williams distribution and the Photon TMD are being probed in the moderate-to-large-$x$ region. In the small-$x$ region the gluon density rises faster with decreasing $x$ than both photon distributions, which behave as $1/x$. Further, in the moderate-to-large-$x$ region, the photon distributions fall much less steeply with increasing $x$ than the gluon density. These two effects combine to give the widely differing behaviour of the cross-section in the backward and forward regions. In the forward region ($x_F,\eta>0$), both the Photon TMD and the Gaussian WW approximation give almost identical results, whereas in the backward region the Photon TMD gives slightly smaller results for moderate values of negative $x_F$. This is similar to what was observed at COMPASS. At fixed $\eta$, the cross-section decreases with increasing $P_T$ by four orders of magnitude in the range $0.5<P_T<3.0$ GeV. With a larger value of the unpolarised TMD width $\langle k_\perp^2\rangle=1.0$ GeV$^2$, the cross-section at fixed $P_T$ is found to be unaffected in the forward region, but shows a decrease in the backward region of about $40\%$ on average. The increase in the $P_T$-spread of the cross-section is also observed at fixed $\eta$, but the effect is very small. The cross-section is also found to be insensitive to changes in the width of the fragmentation function in the range $0<\langle k_{\perp D}^2\rangle<0.25$ GeV$^2$.
\begin{figure}[t]
\begin{center}
\vspace*{-2cm}
\includegraphics[width=\linewidth]{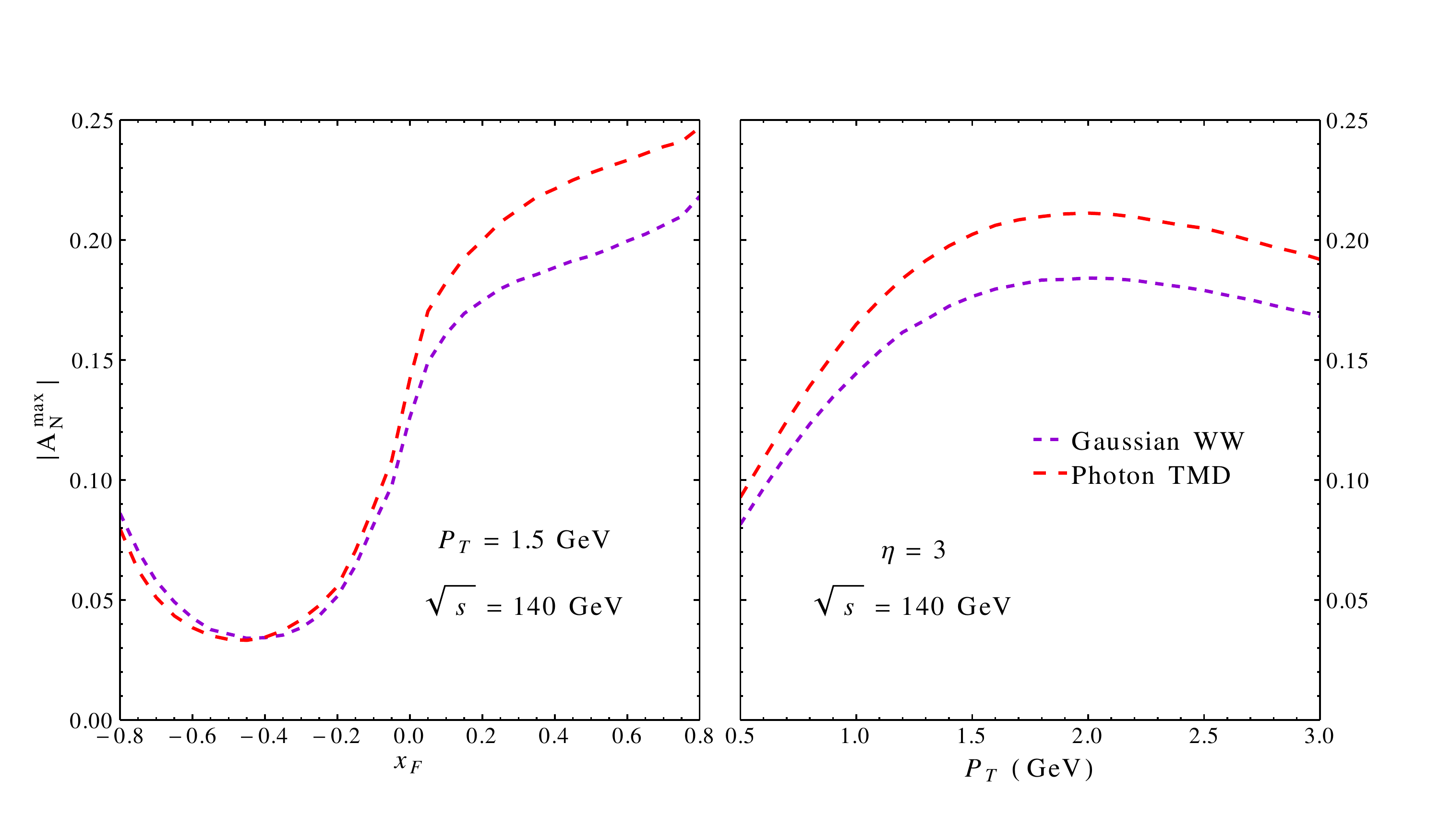}
\vspace*{-2cm}
\caption{SSA with saturated GSF at EIC as a function of $x_F$ (at fixed $P_T$, left panel) and $P_T$ (at fixed $\eta$, right panel).}
\label{EICsatAN}
\end{center}
\end{figure}

In Fig.~\ref{EICsatAN}, we show estimates for the maximal value of the magnitude of the asymmetry $|A_N^\text{max}|$ obtained by using the saturated gluon Sivers function, as a function of $x_F$ at fixed $P_T=1.5$ GeV (left panel) and as a function of $P_T$ at fixed pseudorapidity $\eta=3$ (right panel). With the fairly large centre of mass energy of the EIC, we find that the general features of $|A^\text{max}_N|$ are similar to what was obtained for proton-proton collisions at RHIC~\cite{Anselmino:2004nk,Godbole:2016tvq}.  At fixed $\eta=3$, for the Photon TMD, the asymmetry peaks at $21\%$ at $P_T=2$ GeV. At fixed $P_T$, large asymmetries are allowed in the forward region, with estimates being upto almost 25\% at $x_F=0.8$. Overall, in the forward region ($x_F,\eta>0$) the Photon TMD gives results that are upto 18\% larger than the what is obtained with the Gaussian WW approximation. This difference can be understood qualitatively, from the much smaller values of $k_{\perp\gamma}$ contributing to production in the case of the Photon TMD and the resultant change in the values of $x_g$, $k_{\perp g}$ and $x_\gamma$ which contribute for a given $P_T$ and $x_F$.   As is the case for calculations at RHIC energy and kinematics~\cite{Anselmino:2004nk,Godbole:2016tvq}, the asymmetry is suppressed in the backward hemisphere of the proton ($x_F<0$). This is because, in the backward region, the hard-part $d\sigma/d\that$ depends very weakly on the azimuthal angle of the gluon transverse momentum $\phi_{\perp g}$. This weak dependence, along with the $\cos\phi_{\perp g}$ term that is present in the Sivers function (see Eq.~\ref{Sivers}) leads to a suppression when the azimuthal angle is integrated over. The same has been observed in Ref.~\cite{Anselmino:2004nk}. It must be mentioned however, that this feature is energy dependent and the suppression is weaker at lower centre of mass energies. This can be seen from the large values of $|A^\text{max}_N|$ at $x_F\gtrsim0.3$ for COMPASS shown in Fig.~\ref{COMPsatAN}.

\begin{figure}[t]
\begin{center}
\vspace*{-2cm}
\includegraphics[width=\linewidth]{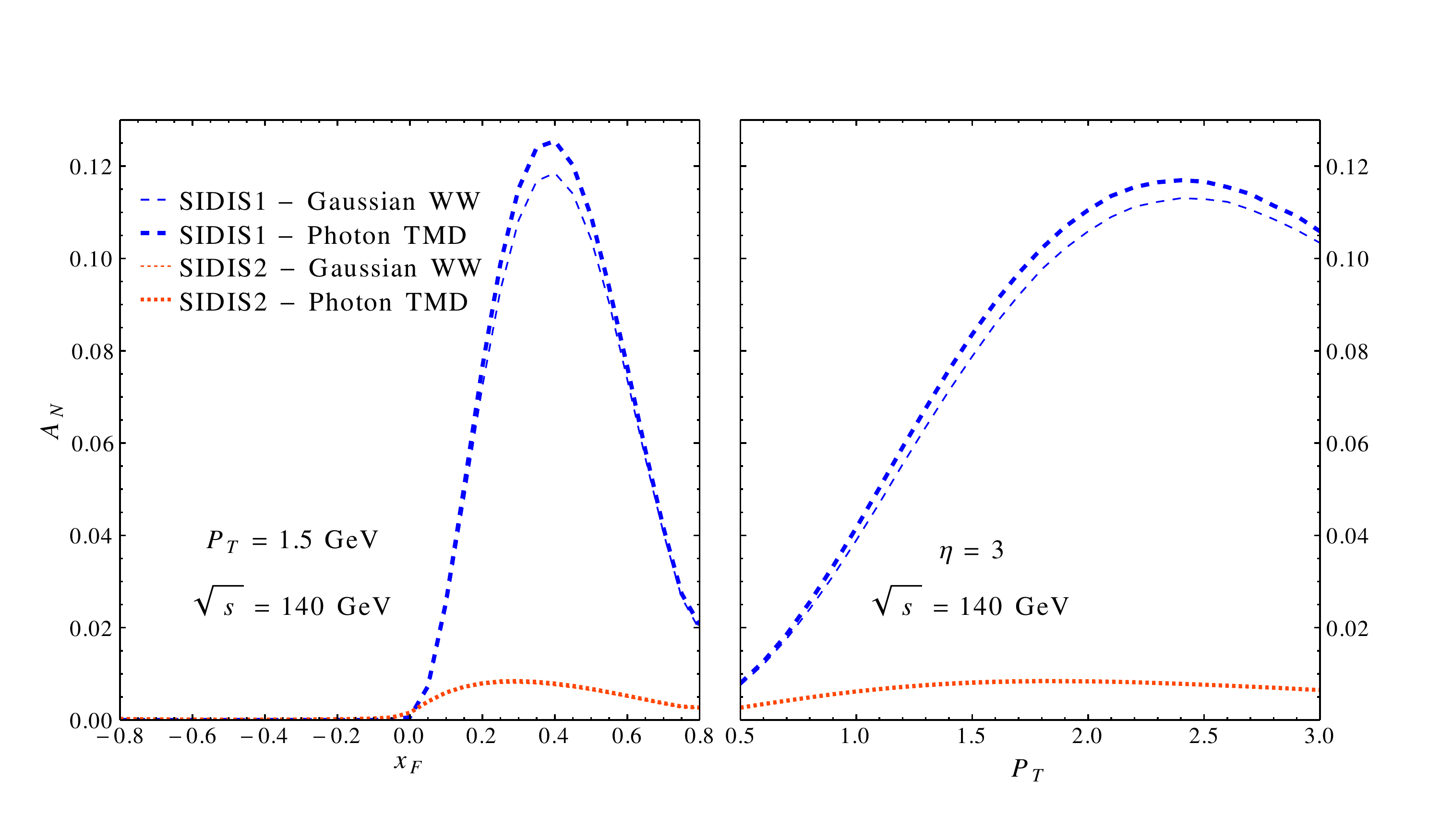}
\vspace*{-2cm}
\caption{SSA from GSF fits of Ref.~\cite{DAlesio:2015fwo} at EIC as a function of $x_F$ (at fixed $P_T$, left panel) and $P_T$ (at fixed $\eta$, right panel).}
\label{EICsidisAN}
\end{center}
\end{figure}

Fig.~\ref{EICsidisAN} shows the asymmetries obtained using the SIDIS1 and SIDIS2 fits~\cite{DAlesio:2015fwo}. As was the case for COMPASS kinematics, both fits give asymmetries much smaller than allowed by the positivity bound, with SIDIS1 giving the larger results of the two. As was found at COMPASS energy, the Photon TMD gives results that are a few percent larger than those obtained using the Gaussian WW approximation.  At fixed $P_T$, in the forward region, SIDIS1 gives a peak asymmetry of 13\% at $x_F=0.4$ and SIDIS2 gives a peak asymmetry of $0.8\%$ at $x_F=0.3$. In the backward region $x_F<0$, $D$ production gets contributions mainly from from $x_g<0.08$, where SIDIS2 is larger than SIDIS1. However the overall values of both fits in this region are very small. Combined with the azimuthal suppression, this makes the asymmetries from both fits almost negligible. At fixed pseudorapidity, SIDIS1 gives a peak asymmetry of around 11.5\% at $P_T=2.5$ GeV and SIDIS2 gives a peak asymmetry of around $0.8\%$ at $P_T=1.6\text{-}2.2$ GeV.

\subsection{Single-Spin Asymmetry in open-charm decay muons}

So far we have considered the SSA in terms of the $D$-meson kinematics. It would also be interesting to consider the SSA in terms of the kinematics of the decay muons. A detector such as the proposed ePHENIX~\cite{Adare:2014aaa} would be able to study open heavy flavour production through the leptonic decay channels. With this in mind, we consider the semileptonic decay of the $D$'s in order to obtain the SSA for the decay muons, ${A^\mu_N}$,
\be
A^\mu_N=\frac{d\sigma^{P^\uparrow l\to D+X\to\mu+X'}-d\sigma^{P^\downarrow l\to D+X\to\mu+X'}}{d\sigma^{P^\uparrow l\to D+X\to\mu+X'}+d\sigma^{P^\downarrow l\to D+X\to\mu+X'}}
\ee
where $d\sigma^{P~l\to D+X\to\mu+X'}$ is the Lorentz-invariant inclusive decay-muon cross-section,
\begin{equation*}
d\sigma^{P~l\to D+X\to\mu+X'} 	\equiv E_\mu \frac{d\sigma^{P~l\to D+X\to\mu+X'}}{d^3\bfp_\mu}.
\end{equation*}
In keeping with the conventions for the $D$-meson asymmetry defined in Eq~1, we take the muon to be produced in the $xz$ plane with the proton moving along the $+z$ direction and its spin parallel or antiparallel to the $y$-axis. Using the narrow width approximation, the expression for the Lorentz-invariant decay-muon cross-section can be written for a general $n$-body decay channel as follows:
\bea
\hspace*{-0.2
cm}
E_\mu \frac{d\sigma^{P~l\to D+X\to\mu+X'}}{d^3\bfp_\mu}&=&\int\frac{d^3\bfp_D}{E_D}\left( E_D\frac{d^3\sigma^{P~l\to D+X}}{d^3\bfp_D}\right)\frac{1}{2(2\pi)^3 E_D~\Gamma^\text{total}}\left(\prod_{i=1}^{n-1}\frac{d^3\bfp_{x_i}}{(2\pi)^32E_{x_i}}\right)\\  \nonumber
&&\hspace*{-4cm}\times|\mathcal{M}^{D^0\to \mu^++x_1+...+x_{n-1}}|^2(2\pi)^4\delta^4\left(P_D-P_\mu-\sum P_{x_i}\right)\times \text{BR}
\label{woo}
\eea
where the $x_i$ are the $n-1$ decay products produced along with the muon. $\Gamma^\text{total}$ is the total decay width of the $D$-meson and BR stands for the branching ratio for the considered $n$-body decay channel. The above expression consists of the meson production cross-section, the decay matrix element and a phase-space integral over all the decay products except the muon. It makes use of the fact that the decay of a scalar meson can be treated as independent of its production, allowing a factorised form involving the meson cross-section convoluted with the differential decay rate. This can be shown to be true by using the narrow width approximation. 

To account for all possible open-charmed meson decays in to muons through all possible channels would be a complex task. To simplify things, we make the following assumptions: First, we consider only the decay of the $D^0$ to muons. Muons can also be produced through the decay of other charmed mesons states but we do not consider those decays here. For the $D^0$ we consider the two major semileptonic decay channels, $D^0\to K^-\mu^+\nu_\mu$ which as a branching ratio of 3.33\% and $D^0\to K^*(892)^-\mu^+\nu_\mu$ which has a branching ratio of 1.92\%. Second, in the calculation of the three-body decays, we set the decay matrix elements $|\mathcal{M}^{D^0\to K^-\mu^+\bar\nu_\mu}|$ and $|\mathcal{M}^{D^0\to K^{*-}\mu^+\bar\nu_\mu}|$ to 1, and only account for the phase-space kinematics. In Eq.~\ref{woo}, the momentum  $\bfp_D$ must be integrated over the entire region of phase-space from which a $D$-meson can decay to produce a muon of given momentum $\bfp_\mu$. The derivation of the closed form expression for decay-muon invariant cross-section, $E_\mu \frac{d\sigma^{P~l\to D+X\to\mu+X'}}{d^3\bfp_\mu}$ and the integration limits for the momentum of the $D$-meson, $\bfp_D$ is given in Appendix B.

\begin{figure}[t]
\begin{center}
\vspace*{-2cm}
\includegraphics[width=1.05\linewidth]{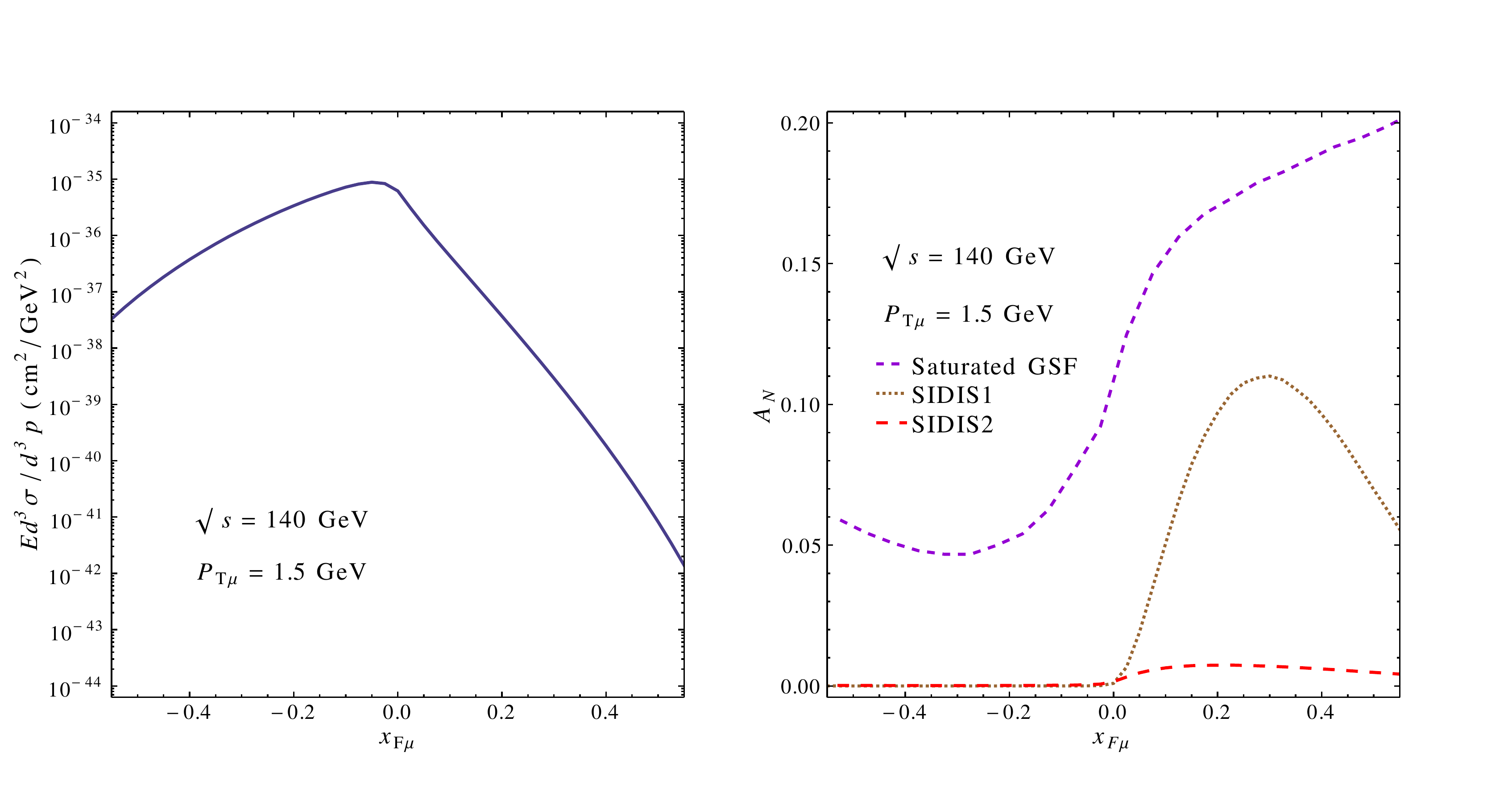}
\vspace*{-2cm}
\caption{Cross-section (left panel) and SSA (right panel) for decay-muons.}
\label{EICsatANdecay}
\end{center}
\end{figure}

The results for the decay-muon invariant cross-section and SSA, $A^\mu_N$ are presented in Fig.~\ref{EICsatANdecay}. The asymmetry is shown for the case of the saturated GSF and the SIDIS1 and SIDIS2~\cite{DAlesio:2015fwo} fits. $A^\mu_N$ is presented as a function of ${x_F}_\mu\equiv 2{P_L}_\mu/\sqrt{s}$, with the muon transverse momentum ${P_T}_\mu=1.5$ GeV. It appears that an azimuthal anisotropy in $D$ production would be retained significantly in the decay-muons. The general dependence of the $A^\mu_N$ on ${x_F}_\mu$ is similar to the dependence of the $D$-meson SSA on $x_F$. As with the $D$-meson,  the muon SSA is also suppressed in the backward hemisphere. Peak values of the $A^\mu_N$ are close to those obtained for the meson: With the Gaussian WW approximation and SIDIS1 GSF, $A^\mu_N$  has a peak value of 11\% at ${x_F}_\mu=0.3$ whereas $A_N$ has a peak value of almost 12\% at $x_F=0.4$. With the SIDIS2 GSF, $A^\mu_N$ has a peak value of 0.75\% at ${x_F}_\mu=0.23$ whereas $A_N$ has a peak value of 0.8\% at $x_F=0.3$.

\section{Conclusions}

In this work, we have presented results for SSA in the low-virtuality leptoproduction of open-charm at both COMPASS and a future Electron-Ion Collider. We find that an asymmetry of upto around 25\% is allowed by the saturated gluon Sivers function at both COMPASS and EIC. We also find that, for EIC kinematics, the asymmetry is significantly retained in the distribution of the decay muons. In calculating the asymmetry we used two different forms for the TMD distribution of quasi-real photons in the lepton. The first was the Weizsacker-Williams distribution with a Gaussian transverse-momentum spread (Gaussian WW) and the second was the LO analytical result for the TMD distribution of photons in a lepton (Photon TMD) from Ref.~\cite{Bacchetta:2015qka}. At COMPASS energy the two forms give similar results, whereas at EIC energy, the Photon TMD gives slightly larger asymmetries in the forward region. The differences in the result for the two distributions can be attributed to the interplay of different $k_{\perp\gamma}$, $x_\gamma$, $x_g$ and and ${k}_{\perp g}$ values that get sampled in the two cases for a given value of $x_F$/$\eta$ and $P_T$.

The two GSF fits of Ref.~\cite{DAlesio:2015fwo} for which we give predictions, were extracted from data on midrapidity pion production at RHIC. As mentioned earlier, the two differ in the flavour structure of the QSFs used as well as the light quark fragmentation functions used in the extraction. SIDIS1 was obtained using an extraction of the QSFs~\cite{Anselmino:2005ea} that included only the $u$ and $d$ flavours and used fragmentation functions by Kretzer~\cite{Kretzer:2000yf}. SIDIS2 was obtained using an extraction of the QSFs that also included sea quarks~\cite{Anselmino:2008sga} and used more recent fragmentation functions by de Florian, Sassot and Stratmann~\cite{deFlorian:2007aj}. While both the GSF fits, taken along with their associated QSF sets, describe the input data on SSA in midrapidity pion production equally well, they have widely differing $x$-dependencies. This indicates that pion production in the midrapidity region at RHIC is only weakly sensitive to the gluon Sivers function.

In this work,  we find that the low-virtuality leptoproduction of open-charm, which probes the gluon content of the proton directly, is able to discriminate well between these two fits. Thus we see that this process offers a good probe of the gluon Sivers function and can be of help in a global extraction of the Sivers function in a generalised parton model framework. In general, at COMPASS and in the forward region of EIC, SIDIS2 gives small, but non-negligible asymmetry predictions on the level of significant fractions of a percent, whereas SIDIS1 predicts larger asymmetries of the order of a few percent.  This indicates that the leptoproduction of open-charm could be a vital probe in constraining the gluon Sivers function, and also in testing the validity of the GPM framework.

\section{Acknowledgements}

R.M.G. wishes to acknowledge support from the Department of Science and
Technology, India under Grant No. SR/S2/JCB-64/2007 under the J.C. Bose Fellowship scheme.  A.M would like to thank the Department of Science and Technology, India for financial support under Grant No.EMR/2014/0000486. A.M would also like to thank the Theory Division, CERN, Switzerland for their kind hospitality.

\section{Appendix}
\subsection{Treatment of GPM kinematics}
In this work, we have considered inclusive single-particle leptoproduction in the low virtuality regime. This allows us to handle the process in terms of a TMD distribution of quasi-real photons in a lepton, not unlike a TMD distribution of partons in a hadron. The treatment of parton kinematics here is thus similar to the treatment of transverse-momentum-dependent parton kinematics for inclusive single-particle hadroproduction, which can be found in quite a few places~\cite{Feynman:1977yr,Contogouris:1978kh} including Ref.~\cite{DAlesio:2004eso}, where heavy meson final states have been considered.

 The momenta of the proton, lepton and the $D$-meson can be written in the $p\text{--}l$ centre of mass frame as,
\be
 P_P=\frac{\sqrt{s}}{2}(1,0,0,1), P_l=\frac{\sqrt{s}}{2}(1,0,0,-1)\text{ and } P_D=(E_D,P_T,0,P_L)
\ee
where the masses of the proton and lepton have been neglected.

The gluon and the quasi-real photon carry light-cone momentum fractions $x_g=P_g^+/P_P^+$, $x_\gamma=P_\gamma^-/P_l^-$ and transverse momenta $\mathbf{k}_g$ and $\mathbf{k}_\gamma$ respectively. Their momenta are given by,
\bea
P_g&=x_g\frac{\sqrt{s}}{2}\left(1+\frac{k_{\perp g}^2}{x_g^2 s},\frac{2k_{\perp g}}{x_g\sqrt{s}}\cos\phi_{\perp g},\frac{2k_{\perp g}}{x_g\sqrt{s}}\sin\phi_{\perp g},1-\frac{k_{\perp g}^2}{x_g^2 s}\right)
\\\nonumber
P_\gamma&=x_\gamma\frac{\sqrt{s}}{2}\left(1+\frac{k_{\perp \gamma}^2}{x_\gamma^2 s},\frac{2k_{\perp \gamma}}{x_\gamma\sqrt{s}}\cos\phi_{\perp\gamma},\frac{2k_{\perp \gamma}}{x_\gamma\sqrt{s}}\sin\phi_{\perp\gamma},-1+\frac{k_{\perp \gamma}^2}{x_\gamma^2 s}\right)
\eea
where $\phi_{\perp g}$ and $\phi_{\perp\gamma}$ are the azimuthal angles of gluon and photon transverse momenta.

The heavy quark is produced through photon-gluon fusion $g\gamma\rightarrow c\bar{c}$ and then fragments into the heavy meson. The momentum of the heavy quark is described by $z$, the light-cone momentum fraction of the heavy meson and $\mathbf{k}_D$, the transverse momentum of the meson with respect to direction of heavy quark. In a choice of coordinates where the heavy quark momentum, $p_c$ is along the $z$-axis, the $D$-meson momentum can be written as
\be
P_D=(E_D,0,0,|\mathbf{p}_D - \mathbf{k}_D|) +  (0,\mathbf{k}_D)
\ee
where the first term on the right is the component along the heavy quark direction and the second term is the component transverse to it. Here, $\mathbf{k}_D$ is simply $(k_{D_x},k_{D_y},0)=(\mathbf{k}_{D_\perp},0)$. In the lab coordinates however, $\mathbf{k}_D$ can have all three components non-zero and is specified as,
\be
\mathbf{k}_D=k_D(\sin\theta\cos\phi,\sin\theta\sin\phi,\cos\theta) \text{, with } |\mathbf{k}_D|=|\mathbf{k}_{D_\perp}|
\ee
and the orthogonality condition $\mathbf{k}_D.\mathbf{p}_c=0$ ensures that $\mathbf{k}_D$ lies in a plane perpendicular to $\mathbf{p}_c$. The light-cone momentum fraction $z$ is given by,
\be
z=\frac{P_D^+}{P_c^+}=\frac{E_D+|\mathbf{p}_D-\mathbf{k}_D|}{E_c+|\mathbf{p}_c|}=\frac{E_D+\sqrt{\mathbf{p}_D^2-\mathbf{k}_D^2}}{E_c+\sqrt{E_c^2-m_c^2}}
\label{light-cone-z}
\ee
This gives us the expression for the energy of the heavy quark,
\be
E_c=\frac{m_c^2+\left((E_D+\sqrt{\mathbf{p}_D^2-\mathbf{k}_D^2})/z\right)^2}{2\left((E_D+\sqrt{\mathbf{p}_D^2-\mathbf{k}_D^2})/z\right)}.
\label{energy-relation}
\ee
The expression for $\mathbf{p}_c$ can be obtained from the fact that it is collinear with $\mathbf{p}_D - \mathbf{k}_D$ and that the unit vector constructed out of both must therefore be equal,
\be
\mathbf{p}_c=\sqrt{E_c^2-m_c^2}\frac{\mathbf{p}_D-\mathbf{k}_D}{|\mathbf{p}_D-\mathbf{k}_D|}.
\label{momentum-relation}
\ee
Eqs. \ref{energy-relation} and \ref{momentum-relation} relate the energy and momentum of the observed $D$-meson with that of the fragmenting heavy quark for given values of $k_D$ and $z$. 

The term $d^3 \mathbf{k}_D \, 
\delta (\mathbf{k}_D \cdot \hat{\mathbf{p}}_c)$ in Eqs. (\ref{denominator}) and (\ref{numerator}) ensures that the $\mathbf{k}_D$ integration is only over momenta transverse to the fragmenting parton:
\be
d^2\mathbf{k}_{D_\perp}=d^3 \mathbf{k}_D \, 
\delta (\mathbf{k}_D \cdot \hat{\mathbf{p}}_c)=dk_D\text{ }k_D\text{ }d\theta \text{ }d\phi\frac{|\mathbf{p}_D-\mathbf{k}_D|}{P_T\sin\phi_1}\left[\delta(\phi-\phi_1)+\delta(\phi-(2\pi-\phi_1))\right]
\ee
where,
\be
\cos \phi_1=\frac{k_D-P_L\cos\theta}{P_T\sin\theta}
\ee
Limits on $k_D$ can be obtained by requiring $|\cos \phi_1|\leq1$,
\be
\text{max}\left[P_L\cos\theta-P_T\sin\theta,0\right]\leq k_D\leq\text{max}\left[P_L\cos\theta+P_T\sin\theta,0\right].
\ee

Furthermore, the inclusion of intrinsic transverse momenta in the kinematics calls for the following constraints: a) the energy of the incoming parton should not be greater than that of its parent particle, $E_{g(\gamma)}\leq E_{p(l)}$ and, b) the energy of the $D$-meson should not be greater than the energy of the heavy quark $E_D\leq E_c$. The first leads to the following bound on the transverse momenta of the incoming partons,
\be
k_{\perp g(\gamma)}<\sqrt{s}~\min[x_{g(\gamma)}, \sqrt{x_{g(\gamma)}(1-x_{g(\gamma)})}].
\ee
The second constraint, $E_D\leq E_c$ is inherently fulfilled by Eq.~\ref{energy-relation}. However, this alone does not ensure that the heavy quark is more energetic than the $D$-meson in the photon-gluon c.o.m frame.  By demanding $E_c>E_D$ in the $\gamma\text{--}g$ c.o.m frame, we get a lower bound on $\shat$,
\be
\shat\geq2P_D.(P_\gamma+P_g).
\ee
In our earlier work on open charm production (Ref.~\cite {Godbole:2016tvq}) we had not implemented this bound in our calculations. We find that the inclusion of this bound significantly improves the convergence of the integral  in the close vincinity of $x_F=0$.

\subsection{Derivation of Lorentz-invariant decay-muon cross-section}
With the decay matrix element set to unity, the three body decay width can be written as follows,
\bea
\Gamma(D\to x_1,x_ 2, \mu)=\frac{1}{2m_D}\frac{1}{(2\pi)^5}\int\frac{d^3\bfp_\mu}{2E_\mu}~\pi\left(\frac{s-m_{x_1}^2}{s}\right)
\label{diff}
\eea
where $s=(p_{x_1}+p_{x_2})^2=(p_D-p_\mu)^2$ is a Lorentz-invariant quantity. Here, $x_1$ and $x_2$ are the decay products produced along with the $\mu$. We take $x_1$ to be $K^-$ or $K^{*-}$ and $x_2$ to be $\bar\nu_\mu$. Here we consider the muon to be massless. Since we know the decay factorises, we can write the production and decay as a convolution,
\be
\sigma^{P~l\to D+X\to\mu+X'}=\int d\sigma^{P~l\to D+X}\ast\frac{d\Gamma}{\Gamma^\text{total}}\times\text{BR}
\ee
where $d\Gamma$ is the decay-width for an infinitesimal muon momentum region $d^3\bfp_\mu$. Then using Eq.~\ref{diff} we finally have,
\be
E_\mu \frac{d\sigma^{P~l\to D+X\to\mu+X'}}{d^3\bfp_\mu}=\int\frac{d^3\bfp_D}{E_D}\left( E_D\frac{d^3\sigma^{P~l\to D+X}}{d^3\bfp_D}\right)\frac{\pi}{4E_D(2\pi)^5}\times\left(\frac{s-m_1^2}{s}\right)\frac{1}{\Gamma^\text{total}}\times\text{BR}.
\ee

The allowed phase-space region for $\bfp_D$ can be obtained determined by considering the decay in the rest frame of the $D$-meson. In it, one can see that the allowed values of muon energy lie in the range $0<E^\text{D com}_\mu<(m_D^2-m_K^2)/2m_D$, where $m_K$ is the mass of the kaon. This constraint on the muon energy in the rest frame of the $D$, can be translated into the following constraint on the the Lorentz-invariant quantity constructed from the $D$-meson and muon four-momenta: 
\be
m_K^2\leq s=(P_D-P_\mu)^2\leq m_D^2
\label{ineq}
\ee
The expression for $s$ in terms of the momenta involved is,
\bea
s=(P_D-P_\mu)^2=(E_D-E_\mu)^2-P_T^2-P_{T\mu}^2+2P_T P_{T\mu} \cos\phi_{P_T}-(P_L-P_{L\mu})^2 
\label{sexp}
\eea
where we have assumed that the muon is massless and is in the $xz$-plane, $\vec{P}_{T\mu}=(P_{T\mu},0)$. Here $P_{T\mu}$ and $P_{L\mu}$ are the $x$ and $z$ components of the muon momentum respectively. The lower inequality in Eq.~\ref{ineq} can be cast as follows:
\be
\cos\phi_{P_T} \geq \frac{2(E_D E_\mu-P_LP_{L\mu})-(m_D^2-m_K^2)}{2P_TP_{T\mu}} (\equiv Y).
\ee
We will call the quantity on the right hand side of the above expression, $Y$. This gives us a constraint on the angle of the $D$-meson transverse momentum,
\be
-\cos^{-1}Y\leq\phi_{P_T}\leq\cos^{-1}Y
\ee
Naturally, we also require $Y\leq 1$, since the lower bound on a cosine term can't be greater than 1. Demanding this gives us upper and lower limits on $P_L$:
\bea
P_L^\text{max}&=&\frac{1}{2P_{T\mu}^2}[P_{L\mu}(m_D^2-m_K^2+2P_TP_{T\mu})  \nonumber \\
&+&
\sqrt{
E_\mu^2 \left(
(m_D^2-m_K^2)^2+2P_{T\mu}(P_T(m_D^2-2m_K^2)-m_D^2P_{T\mu})
\right)
}
]
\eea
\bea
P_L^\text{min}&=&\frac{1}{2P_{T\mu}^2}[P_{L\mu}(m_D^2-m_K^2+2P_TP_{T\mu})  \nonumber \\
&-&
\sqrt{
E_\mu^2 \left(
(m_D^2-m_K^2)^2+2P_{T\mu}(P_T(m_D^2-2m_K^2)-m_D^2P_{T\mu})
\right)
}
]
\eea
Demanding $Y\leq 1$ also gives us a lower bound on $P_T$:
\be
P_T^\text{min}=\max[0,\frac{4m_D^2P_{T\mu}^2-(m_D^2-m_K^2)^2}{4(m_D^2-m_K^2)P_{T\mu}}]
\ee


\begin{thebibliography}{0}
\expandafter\ifx\csname natexlab\endcsname\relax\def\natexlab#1{#1}\fi
\expandafter\ifx\csname bibnamefont\endcsname\relax
  \def\bibnamefont#1{#1}\fi
\expandafter\ifx\csname bibfnamefont\endcsname\relax
  \def\bibfnamefont#1{#1}\fi
\expandafter\ifx\csname citenamefont\endcsname\relax
  \def\citenamefont#1{#1}\fi
\expandafter\ifx\csname url\endcsname\relax
  \def\url#1{\texttt{#1}}\fi
\expandafter\ifx\csname urlprefix\endcsname\relax\def\urlprefix{URL }\fi
\providecommand{\bibinfo}[2]{#2}
\providecommand{\eprint}[2][]{\url{#2}}

\end{thebibliography}


\begin{thebibliography}{33}
\expandafter\ifx\csname natexlab\endcsname\relax\def\natexlab#1{#1}\fi
\expandafter\ifx\csname bibnamefont\endcsname\relax
  \def\bibnamefont#1{#1}\fi
\expandafter\ifx\csname bibfnamefont\endcsname\relax
  \def\bibfnamefont#1{#1}\fi
\expandafter\ifx\csname citenamefont\endcsname\relax
  \def\citenamefont#1{#1}\fi
\expandafter\ifx\csname url\endcsname\relax
  \def\url#1{\texttt{#1}}\fi
\expandafter\ifx\csname urlprefix\endcsname\relax\def\urlprefix{URL }\fi
\providecommand{\bibinfo}[2]{#2}
\providecommand{\eprint}[2][]{\url{#2}}

\bibitem[{\citenamefont{Dick et~al.}(1975)}]{Dick:1975ty}
\bibinfo{author}{\bibfnamefont{L.}~\bibnamefont{Dick}} \bibnamefont{et~al.},
  \bibinfo{journal}{Phys. Lett.} \textbf{\bibinfo{volume}{57B}},
  \bibinfo{pages}{93} (\bibinfo{year}{1975}).

\bibitem[{\citenamefont{Klem et~al.}(1976)\citenamefont{Klem, Bowers, Courant,
  Kagan, Marshak, Peterson, Ruddick, Dragoset, and Roberts}}]{Klem:1976ui}
\bibinfo{author}{\bibfnamefont{R.~D.} \bibnamefont{Klem}},
  \bibinfo{author}{\bibfnamefont{J.~E.} \bibnamefont{Bowers}},
  \bibinfo{author}{\bibfnamefont{H.~W.} \bibnamefont{Courant}},
  \bibinfo{author}{\bibfnamefont{H.}~\bibnamefont{Kagan}},
  \bibinfo{author}{\bibfnamefont{M.~L.} \bibnamefont{Marshak}},
  \bibinfo{author}{\bibfnamefont{E.~A.} \bibnamefont{Peterson}},
  \bibinfo{author}{\bibfnamefont{K.}~\bibnamefont{Ruddick}},
  \bibinfo{author}{\bibfnamefont{W.~H.} \bibnamefont{Dragoset}},
  \bibnamefont{and} \bibinfo{author}{\bibfnamefont{J.~B.}
  \bibnamefont{Roberts}}, \bibinfo{journal}{Phys. Rev. Lett.}
  \textbf{\bibinfo{volume}{36}}, \bibinfo{pages}{929} (\bibinfo{year}{1976}).

\bibitem[{\citenamefont{Dragoset et~al.}(1978)\citenamefont{Dragoset, Roberts,
  Bowers, Courant, Kagan, Marshak, Peterson, Ruddick, and
  Klem}}]{Dragoset:1978gg}
\bibinfo{author}{\bibfnamefont{W.~H.} \bibnamefont{Dragoset}},
  \bibinfo{author}{\bibfnamefont{J.~B.} \bibnamefont{Roberts}},
  \bibinfo{author}{\bibfnamefont{J.~E.} \bibnamefont{Bowers}},
  \bibinfo{author}{\bibfnamefont{H.~W.} \bibnamefont{Courant}},
  \bibinfo{author}{\bibfnamefont{H.}~\bibnamefont{Kagan}},
  \bibinfo{author}{\bibfnamefont{M.~L.} \bibnamefont{Marshak}},
  \bibinfo{author}{\bibfnamefont{E.~A.} \bibnamefont{Peterson}},
  \bibinfo{author}{\bibfnamefont{K.}~\bibnamefont{Ruddick}}, \bibnamefont{and}
  \bibinfo{author}{\bibfnamefont{R.~D.} \bibnamefont{Klem}},
  \bibinfo{journal}{Phys. Rev.} \textbf{\bibinfo{volume}{D18}},
  \bibinfo{pages}{3939} (\bibinfo{year}{1978}).

\bibitem[{\citenamefont{D'Alesio and Murgia}(2008)}]{DAlesio:2007bjf}
\bibinfo{author}{\bibfnamefont{U.}~\bibnamefont{D'Alesio}} \bibnamefont{and}
  \bibinfo{author}{\bibfnamefont{F.}~\bibnamefont{Murgia}},
  \bibinfo{journal}{Prog. Part. Nucl. Phys.} \textbf{\bibinfo{volume}{61}},
  \bibinfo{pages}{394} (\bibinfo{year}{2008}), \eprint{0712.4328}.

\bibitem[{\citenamefont{Barone et~al.}(2010)\citenamefont{Barone, Bradamante,
  and Martin}}]{Barone:2010zz}
\bibinfo{author}{\bibfnamefont{V.}~\bibnamefont{Barone}},
  \bibinfo{author}{\bibfnamefont{F.}~\bibnamefont{Bradamante}},
  \bibnamefont{and} \bibinfo{author}{\bibfnamefont{A.}~\bibnamefont{Martin}},
  \bibinfo{journal}{Prog. Part. Nucl. Phys.} \textbf{\bibinfo{volume}{65}},
  \bibinfo{pages}{267} (\bibinfo{year}{2010}), \eprint{1011.0909}.

\bibitem[{\citenamefont{D'Alesio and Murgia}(2004)}]{DAlesio:2004eso}
\bibinfo{author}{\bibfnamefont{U.}~\bibnamefont{D'Alesio}} \bibnamefont{and}
  \bibinfo{author}{\bibfnamefont{F.}~\bibnamefont{Murgia}},
  \bibinfo{journal}{Phys. Rev.} \textbf{\bibinfo{volume}{D70}},
  \bibinfo{pages}{074009} (\bibinfo{year}{2004}), \eprint{hep-ph/0408092}.

\bibitem[{\citenamefont{Sivers}(1990)}]{Sivers:1989cc}
\bibinfo{author}{\bibfnamefont{D.~W.} \bibnamefont{Sivers}},
  \bibinfo{journal}{Phys. Rev.} \textbf{\bibinfo{volume}{D41}},
  \bibinfo{pages}{83} (\bibinfo{year}{1990}).

\bibitem[{\citenamefont{Sivers}(1991)}]{Sivers:1990fh}
\bibinfo{author}{\bibfnamefont{D.~W.} \bibnamefont{Sivers}},
  \bibinfo{journal}{Phys. Rev.} \textbf{\bibinfo{volume}{D43}},
  \bibinfo{pages}{261} (\bibinfo{year}{1991}).

\bibitem[{\citenamefont{D'Alesio et~al.}(2015)\citenamefont{D'Alesio, Murgia,
  and Pisano}}]{DAlesio:2015fwo}
\bibinfo{author}{\bibfnamefont{U.}~\bibnamefont{D'Alesio}},
  \bibinfo{author}{\bibfnamefont{F.}~\bibnamefont{Murgia}}, \bibnamefont{and}
  \bibinfo{author}{\bibfnamefont{C.}~\bibnamefont{Pisano}},
  \bibinfo{journal}{JHEP} \textbf{\bibinfo{volume}{09}}, \bibinfo{pages}{119}
  (\bibinfo{year}{2015}), \eprint{1506.03078}.

\bibitem[{\citenamefont{Adolph et~al.}(2017)}]{Adolph:2017pgv}
\bibinfo{author}{\bibfnamefont{C.}~\bibnamefont{Adolph}} \bibnamefont{et~al.}
  (\bibinfo{collaboration}{COMPASS}), \bibinfo{journal}{Phys. Lett.}
  \textbf{\bibinfo{volume}{B772}}, \bibinfo{pages}{854} (\bibinfo{year}{2017}),
  \eprint{1701.02453}.

\bibitem[{\citenamefont{Anselmino et~al.}(2004)\citenamefont{Anselmino,
  Boglione, D'Alesio, Leader, and Murgia}}]{Anselmino:2004nk}
\bibinfo{author}{\bibfnamefont{M.}~\bibnamefont{Anselmino}},
  \bibinfo{author}{\bibfnamefont{M.}~\bibnamefont{Boglione}},
  \bibinfo{author}{\bibfnamefont{U.}~\bibnamefont{D'Alesio}},
  \bibinfo{author}{\bibfnamefont{E.}~\bibnamefont{Leader}}, \bibnamefont{and}
  \bibinfo{author}{\bibfnamefont{F.}~\bibnamefont{Murgia}},
  \bibinfo{journal}{Phys. Rev.} \textbf{\bibinfo{volume}{D70}},
  \bibinfo{pages}{074025} (\bibinfo{year}{2004}), \eprint{hep-ph/0407100}.

\bibitem[{\citenamefont{Godbole et~al.}(2016)\citenamefont{Godbole, Kaushik,
  and Misra}}]{Godbole:2016tvq}
\bibinfo{author}{\bibfnamefont{R.~M.} \bibnamefont{Godbole}},
  \bibinfo{author}{\bibfnamefont{A.}~\bibnamefont{Kaushik}}, \bibnamefont{and}
  \bibinfo{author}{\bibfnamefont{A.}~\bibnamefont{Misra}},
  \bibinfo{journal}{Phys. Rev.} \textbf{\bibinfo{volume}{D94}},
  \bibinfo{pages}{114022} (\bibinfo{year}{2016}), \eprint{1606.01818}.

\bibitem[{\citenamefont{Godbole et~al.}(2012)\citenamefont{Godbole, Misra,
  Mukherjee, and Rawoot}}]{Godbole:2012bx}
\bibinfo{author}{\bibfnamefont{R.~M.} \bibnamefont{Godbole}},
  \bibinfo{author}{\bibfnamefont{A.}~\bibnamefont{Misra}},
  \bibinfo{author}{\bibfnamefont{A.}~\bibnamefont{Mukherjee}},
  \bibnamefont{and} \bibinfo{author}{\bibfnamefont{V.~S.}
  \bibnamefont{Rawoot}}, \bibinfo{journal}{Phys. Rev.}
  \textbf{\bibinfo{volume}{D85}}, \bibinfo{pages}{094013}
  (\bibinfo{year}{2012}), \eprint{1201.1066}.

\bibitem[{\citenamefont{Godbole et~al.}(2013)\citenamefont{Godbole, Misra,
  Mukherjee, and Rawoot}}]{Godbole:2013bca}
\bibinfo{author}{\bibfnamefont{R.~M.} \bibnamefont{Godbole}},
  \bibinfo{author}{\bibfnamefont{A.}~\bibnamefont{Misra}},
  \bibinfo{author}{\bibfnamefont{A.}~\bibnamefont{Mukherjee}},
  \bibnamefont{and} \bibinfo{author}{\bibfnamefont{V.~S.}
  \bibnamefont{Rawoot}}, \bibinfo{journal}{Phys. Rev.}
  \textbf{\bibinfo{volume}{D88}}, \bibinfo{pages}{014029}
  (\bibinfo{year}{2013}), \eprint{1304.2584}.

\bibitem[{\citenamefont{Godbole et~al.}(2015)\citenamefont{Godbole, Kaushik,
  Misra, and Rawoot}}]{Godbole:2014tha}
\bibinfo{author}{\bibfnamefont{R.~M.} \bibnamefont{Godbole}},
  \bibinfo{author}{\bibfnamefont{A.}~\bibnamefont{Kaushik}},
  \bibinfo{author}{\bibfnamefont{A.}~\bibnamefont{Misra}}, \bibnamefont{and}
  \bibinfo{author}{\bibfnamefont{V.~S.} \bibnamefont{Rawoot}},
  \bibinfo{journal}{Phys. Rev.} \textbf{\bibinfo{volume}{D91}},
  \bibinfo{pages}{014005} (\bibinfo{year}{2015}), \eprint{1405.3560}.

\bibitem[{\citenamefont{D'Alesio
  et~al.}(2017{\natexlab{a}})\citenamefont{D'Alesio, Flore, and
  Murgia}}]{DAlesio:2017nrd}
\bibinfo{author}{\bibfnamefont{U.}~\bibnamefont{D'Alesio}},
  \bibinfo{author}{\bibfnamefont{C.}~\bibnamefont{Flore}}, \bibnamefont{and}
  \bibinfo{author}{\bibfnamefont{F.}~\bibnamefont{Murgia}},
  \bibinfo{journal}{Phys. Rev.} \textbf{\bibinfo{volume}{D95}},
  \bibinfo{pages}{094002} (\bibinfo{year}{2017}{\natexlab{a}}),
  \eprint{1701.01148}.

\bibitem[{\citenamefont{D'Alesio
  et~al.}(2017{\natexlab{b}})\citenamefont{D'Alesio, Flore, and
  Murgia}}]{DAlesio:2017jdm}
\bibinfo{author}{\bibfnamefont{U.}~\bibnamefont{D'Alesio}},
  \bibinfo{author}{\bibfnamefont{C.}~\bibnamefont{Flore}}, \bibnamefont{and}
  \bibinfo{author}{\bibfnamefont{F.}~\bibnamefont{Murgia}},
  \bibinfo{journal}{PoS} \textbf{\bibinfo{volume}{QCDEV2016}},
  \bibinfo{pages}{002} (\bibinfo{year}{2017}{\natexlab{b}}),
  \eprint{1701.03303}.

\bibitem[{\citenamefont{Yuan}(2008)}]{Yuan:2008vn}
\bibinfo{author}{\bibfnamefont{F.}~\bibnamefont{Yuan}}, \bibinfo{journal}{Phys.
  Rev.} \textbf{\bibinfo{volume}{D78}}, \bibinfo{pages}{014024}
  (\bibinfo{year}{2008}), \eprint{0801.4357}.

\bibitem[{\citenamefont{Accardi et~al.}(2016)}]{Accardi:2012qut}
\bibinfo{author}{\bibfnamefont{A.}~\bibnamefont{Accardi}} \bibnamefont{et~al.},
  \bibinfo{journal}{Eur. Phys. J.} \textbf{\bibinfo{volume}{A52}},
  \bibinfo{pages}{268} (\bibinfo{year}{2016}), \eprint{1212.1701}.

\bibitem[{\citenamefont{Babcock et~al.}(1978)\citenamefont{Babcock, Sivers, and
  Wolfram}}]{Babcock:1977fi}
\bibinfo{author}{\bibfnamefont{J.}~\bibnamefont{Babcock}},
  \bibinfo{author}{\bibfnamefont{D.~W.} \bibnamefont{Sivers}},
  \bibnamefont{and} \bibinfo{author}{\bibfnamefont{S.}~\bibnamefont{Wolfram}},
  \bibinfo{journal}{Phys. Rev.} \textbf{\bibinfo{volume}{D18}},
  \bibinfo{pages}{162} (\bibinfo{year}{1978}).

\bibitem[{\citenamefont{Brodsky et~al.}(1971)\citenamefont{Brodsky, Kinoshita,
  and Terazawa}}]{Brodsky:1971ud}
\bibinfo{author}{\bibfnamefont{S.~J.} \bibnamefont{Brodsky}},
  \bibinfo{author}{\bibfnamefont{T.}~\bibnamefont{Kinoshita}},
  \bibnamefont{and} \bibinfo{author}{\bibfnamefont{H.}~\bibnamefont{Terazawa}},
  \bibinfo{journal}{Phys. Rev.} \textbf{\bibinfo{volume}{D4}},
  \bibinfo{pages}{1532} (\bibinfo{year}{1971}).

\bibitem[{\citenamefont{Terazawa}(1973)}]{Terazawa:1973tb}
\bibinfo{author}{\bibfnamefont{H.}~\bibnamefont{Terazawa}},
  \bibinfo{journal}{Rev. Mod. Phys.} \textbf{\bibinfo{volume}{45}},
  \bibinfo{pages}{615} (\bibinfo{year}{1973}).

\bibitem[{\citenamefont{Kniehl}(1991)}]{Kniehl:1990iv}
\bibinfo{author}{\bibfnamefont{B.~A.} \bibnamefont{Kniehl}},
  \bibinfo{journal}{Phys. Lett.} \textbf{\bibinfo{volume}{B254}},
  \bibinfo{pages}{267} (\bibinfo{year}{1991}).

\bibitem[{\citenamefont{Bacchetta et~al.}(2016)\citenamefont{Bacchetta,
  Mantovani, and Pasquini}}]{Bacchetta:2015qka}
\bibinfo{author}{\bibfnamefont{A.}~\bibnamefont{Bacchetta}},
  \bibinfo{author}{\bibfnamefont{L.}~\bibnamefont{Mantovani}},
  \bibnamefont{and} \bibinfo{author}{\bibfnamefont{B.}~\bibnamefont{Pasquini}},
  \bibinfo{journal}{Phys. Rev.} \textbf{\bibinfo{volume}{D93}},
  \bibinfo{pages}{013005} (\bibinfo{year}{2016}), \eprint{1508.06964}.

\bibitem[{\citenamefont{D'Alesio et~al.}(2011)\citenamefont{D'Alesio, Murgia,
  and Pisano}}]{DAlesio:2010sag}
\bibinfo{author}{\bibfnamefont{U.}~\bibnamefont{D'Alesio}},
  \bibinfo{author}{\bibfnamefont{F.}~\bibnamefont{Murgia}}, \bibnamefont{and}
  \bibinfo{author}{\bibfnamefont{C.}~\bibnamefont{Pisano}},
  \bibinfo{journal}{Phys. Rev.} \textbf{\bibinfo{volume}{D83}},
  \bibinfo{pages}{034021} (\bibinfo{year}{2011}), \eprint{1011.2692}.

\bibitem[{\citenamefont{Anselmino et~al.}(2005)\citenamefont{Anselmino,
  Boglione, D'Alesio, Kotzinian, Murgia, and Prokudin}}]{Anselmino:2005ea}
\bibinfo{author}{\bibfnamefont{M.}~\bibnamefont{Anselmino}},
  \bibinfo{author}{\bibfnamefont{M.}~\bibnamefont{Boglione}},
  \bibinfo{author}{\bibfnamefont{U.}~\bibnamefont{D'Alesio}},
  \bibinfo{author}{\bibfnamefont{A.}~\bibnamefont{Kotzinian}},
  \bibinfo{author}{\bibfnamefont{F.}~\bibnamefont{Murgia}}, \bibnamefont{and}
  \bibinfo{author}{\bibfnamefont{A.}~\bibnamefont{Prokudin}},
  \bibinfo{journal}{Phys. Rev.} \textbf{\bibinfo{volume}{D72}},
  \bibinfo{pages}{094007} (\bibinfo{year}{2005}), \bibinfo{note}{[Erratum:
  Phys. Rev.D72,099903(E) (2005)]}, \eprint{hep-ph/0507181}.

\bibitem[{\citenamefont{Kretzer}(2000)}]{Kretzer:2000yf}
\bibinfo{author}{\bibfnamefont{S.}~\bibnamefont{Kretzer}},
  \bibinfo{journal}{Phys. Rev.} \textbf{\bibinfo{volume}{D62}},
  \bibinfo{pages}{054001} (\bibinfo{year}{2000}), \eprint{hep-ph/0003177}.

\bibitem[{\citenamefont{Anselmino et~al.}(2009)\citenamefont{Anselmino,
  Boglione, D'Alesio, Kotzinian, Melis, Murgia, Prokudin, and
  Turk}}]{Anselmino:2008sga}
\bibinfo{author}{\bibfnamefont{M.}~\bibnamefont{Anselmino}},
  \bibinfo{author}{\bibfnamefont{M.}~\bibnamefont{Boglione}},
  \bibinfo{author}{\bibfnamefont{U.}~\bibnamefont{D'Alesio}},
  \bibinfo{author}{\bibfnamefont{A.}~\bibnamefont{Kotzinian}},
  \bibinfo{author}{\bibfnamefont{S.}~\bibnamefont{Melis}},
  \bibinfo{author}{\bibfnamefont{F.}~\bibnamefont{Murgia}},
  \bibinfo{author}{\bibfnamefont{A.}~\bibnamefont{Prokudin}}, \bibnamefont{and}
  \bibinfo{author}{\bibfnamefont{C.}~\bibnamefont{Turk}},
  \bibinfo{journal}{Eur. Phys. J.} \textbf{\bibinfo{volume}{A39}},
  \bibinfo{pages}{89} (\bibinfo{year}{2009}), \eprint{0805.2677}.

\bibitem[{\citenamefont{de~Florian et~al.}(2007)\citenamefont{de~Florian,
  Sassot, and Stratmann}}]{deFlorian:2007aj}
\bibinfo{author}{\bibfnamefont{D.}~\bibnamefont{de~Florian}},
  \bibinfo{author}{\bibfnamefont{R.}~\bibnamefont{Sassot}}, \bibnamefont{and}
  \bibinfo{author}{\bibfnamefont{M.}~\bibnamefont{Stratmann}},
  \bibinfo{journal}{Phys. Rev.} \textbf{\bibinfo{volume}{D75}},
  \bibinfo{pages}{114010} (\bibinfo{year}{2007}), \eprint{hep-ph/0703242}.

\bibitem[{\citenamefont{Kniehl and Kramer}(2006)}]{Kniehl:2006mw}
\bibinfo{author}{\bibfnamefont{B.~A.} \bibnamefont{Kniehl}} \bibnamefont{and}
  \bibinfo{author}{\bibfnamefont{G.}~\bibnamefont{Kramer}},
  \bibinfo{journal}{Phys. Rev.} \textbf{\bibinfo{volume}{D74}},
  \bibinfo{pages}{037502} (\bibinfo{year}{2006}), \eprint{hep-ph/0607306}.

\bibitem[{\citenamefont{Adare et~al.}(2014)}]{Adare:2014aaa}
\bibinfo{author}{\bibfnamefont{A.}~\bibnamefont{Adare}} \bibnamefont{et~al.}
  (\bibinfo{collaboration}{PHENIX}) (\bibinfo{year}{2014}), \eprint{1402.1209}.

\bibitem[{\citenamefont{Feynman et~al.}(1977)\citenamefont{Feynman, Field, and
  Fox}}]{Feynman:1977yr}
\bibinfo{author}{\bibfnamefont{R.~P.} \bibnamefont{Feynman}},
  \bibinfo{author}{\bibfnamefont{R.~D.} \bibnamefont{Field}}, \bibnamefont{and}
  \bibinfo{author}{\bibfnamefont{G.~C.} \bibnamefont{Fox}},
  \bibinfo{journal}{Nucl. Phys.} \textbf{\bibinfo{volume}{B128}},
  \bibinfo{pages}{1} (\bibinfo{year}{1977}).

\bibitem[{\citenamefont{Contogouris et~al.}(1978)\citenamefont{Contogouris,
  Gaskell, and Papadopoulos}}]{Contogouris:1978kh}
\bibinfo{author}{\bibfnamefont{A.~P.} \bibnamefont{Contogouris}},
  \bibinfo{author}{\bibfnamefont{R.}~\bibnamefont{Gaskell}}, \bibnamefont{and}
  \bibinfo{author}{\bibfnamefont{S.}~\bibnamefont{Papadopoulos}},
  \bibinfo{journal}{Phys. Rev.} \textbf{\bibinfo{volume}{D17}},
  \bibinfo{pages}{2314} (\bibinfo{year}{1978}).

\end{thebibliography}
\end{document}